\documentclass[prd,aps,showpacs,nofootinbib,tightenlines]{revtex4}
\usepackage{xspace}
\usepackage{mathrsfs}
\usepackage{amsmath}
\usepackage{amssymb}
\usepackage{epsfig}
\usepackage{graphicx}
\usepackage{booktabs}
\usepackage{multirow}
\usepackage{subfigure}
\usepackage{bm}
\usepackage{times}
\usepackage{braket}
\usepackage{color}
\usepackage{slashed}
\usepackage{hyperref}
\usepackage{threeparttable}
\definecolor{Red}{rgb}{1.,0.,0.}

\definecolor{Blue}{rgb}{0.,0.,1.}

\definecolor{nicered}{rgb}{0.7,0.1,0.1}
\definecolor{nicegreen}{rgb}{0.1,0.5,0.1}
\hypersetup{colorlinks,citecolor=nicegreen,linkcolor=nicered}
\newcommand{\optbar}[1]{\shortstack{{\tiny (\rule[.4ex]{1em}{.1mm})}\\ [-.7ex] $#1$}}
\newcommand{\beq}{\begin{eqnarray}}
\newcommand{\eeq}{\end{eqnarray}}
\newcommand{\non}{\nonumber\\ }

\newcommand{\KorKbar}{\kern 0.18em\optbar{\kern -0.18em K}{}\xspace}

\def \epjc{ Eur. Phys. J. C }

\def \jhep{ J. High Energy Phys. }

\begin{document}
%XXXXXXXXXXXXXXXXXXXXXXXXXXXXXXXXXX% @ Begin
\title{Global determination of two-meson distribution amplitudes from three-body $B$
decays in the perturbative QCD approach}
\author{Ya Li$^1$}                \email[]{liyakelly@163.com}
\author{Da-Cheng Yan$^2$}          \email[]{yandac@126.com}
\author{Jun Hua$^3$}          \email[]{huajun$\underline{\;}$phy@sjtu.edu.cn}
\author{Zhou Rui$^{4}$}             \email[]{jindui1127@126.com}
\author{Hsiang-nan Li$^5$}         \email[]{hnli@phys.sinica.edu.tw}

\affiliation{$^1$ Department of Physics, College of Sciences, Nanjing Agricultural University,
Nanjing, Jiangsu 210095, China}
\affiliation{$^2$ School of Mathematics and Physics, Changzhou University, Changzhou, Jiangsu 213164, China}
\affiliation{$^3$ INPAC, Key Laboratory for Particle Astrophysics and Cosmology (MOE),
Shanghai Key Laboratory for Particle Physics and Cosmology, School of Physics and Astronomy,
Shanghai Jiao-Tong University, Shanghai 200240, China}
\affiliation{$^4$ College of Sciences, North China University of Science and Technology,
Tangshan, Hebei 063210, China}
\affiliation{$^5$ Institute of Physics, Academia Sinica, Taipei, Taiwan 115, Republic of China}
\date{\today}

%XXXXXXXXXXXXXXXXXXXXXXXXXXXXXXXXXX%
\begin{abstract}
We perform a global analysis of three-body charmless hadronic decays
$B\to VP_3\to P_1P_2P_3$ in the perturbative QCD (PQCD) approach, where $V$ denotes
an intermediate vector resonance, and $P_i$, $i=1,2,3$, denote final-state pseudoscalar mesons.
Fitting the PQCD factorization formulas at leading order in the strong coupling $\alpha_s$
to measured branching ratios and direct $CP$ asymmetries, we determine the
Gegenbauer moments in the two-meson distribution amplitudes (DAs) for the meson pairs
$P_1P_2=\pi\pi, K\pi, KK$. The fitted Gegenbauer moments are then employed to make
predictions for those observables, whose data are excluded in the fit due to
larger experimental uncertainties. A general consistency between our predictions and
data is achieved, which hints the validity of the PQCD formalism for the above
three-body $B$ meson decays and the universality of the nonperturbative two-meson
DAs. The obtained two-meson DAs can be applied to PQCD studies
of other multi-body $B$ meson decays involving the same meson pairs.
We also attempt to determine the dependence of the Gegenbauer moments
on the meson-pair invariant mass, and demonstrate that this determination is promising,
when data become more precise.
\end{abstract}

\maketitle

%XXXXXXXXXXXXXXXXXXXXXXXXXXXXXXXXXX%

%XXXXXXXXXXXXXXXXXXXXXXXXXXXXXXXXXX% @ Begin
\section{Introduction}

Since the perturbative QCD (PQCD) framework for three-body $B$ meson decays was proposed in~\cite{Chen:2002th},
there have been extensive applications to various
channels~\cite{epjc77199,prd99093007,epjc79792,cpc43073103,epjc80394,epjc80517,epjc8191,prd103013005,
prd95056008,epjc7937,prd103016002,epjc80815,prd101111901,jpg46095001}, and rich phenomenology has been
explored.  This formalism is based on the $k_T$ factorization theorem for leading-power regions of
a Dalitz plot, where two final-state mesons are roughly collimated to each other.
The dominant nonperturbative dynamics responsible for the production of the meson pair,
including final-state interactions between the two mesons, is absorbed into two-meson
distribution amplitudes (DAs)~\cite{G,G1,DM,Diehl:1998dk,Diehl:1998dk1,Diehl:1998dk2,MP}.
It is similar to the absorption of collinear divergences associated with a meson,
which participates a high-energy QCD exclusive process, into its meson DAs.
The remaining contributions, being calculable at the parton level in perturbation
theory, go into hard kernels. The analysis of three-body $B$ meson decays is then simplified
to that of two-body decays, where a Feynman diagram for hard kernels at leading order (LO)
of the strong coupling $\alpha_s$ involves a single virtual gluon exchange. The same idea has been
extended to four-body charmless hadronic $B$ meson decays recently~\cite{Rui:2021kbn}: they are
assumed to proceed dominantly with two intermediate resonances, which then strongly decay into two
light meson pairs. Various asymmetries in final-state angular distributions from the
$B_{(s)} \to (K\pi) (K\pi)$ decays were predicted based on the universality of the two-meson DAs
for the $K\pi$ pair.

A two-meson DA, being the time-like version of a generalized parton distribution
function, depends on the parton momentum fraction $x$, the meson momentum fraction $\zeta$,
which describes the relative motion between the two mesons in the pair, and the meson-pair
invariant mass squared $\omega^2$.
For the $x$ dependence, one can decompose a two-meson DA into the eigenfunctions of its
evolution equation~\cite{JETP25-510,plb87-359,plb94-245,JETP26-594},
i.e., the Gegenbauer polynomials $C_n^{3/2}(2x-1)$, based on the conformal
symmetry. This expansion follows that for a hadron DA exactly. As to the expansion in $\zeta$, one employs the
partial waves for the produced meson pair, i.e, the Legendre polynomials $P_l(2\zeta-1)$, noticing
the relation $2\zeta-1=\cos\theta$ with $\theta$ being the polar angle of a meson in the center-of-mass frame
of the meson pair~\cite{Hambrock:2015aor}. The expansion of a two-meson DA in terms of
the two sets of orthogonal polynomials then reads~\cite{MP}
\begin{eqnarray}\label{2mda}
\Phi(x,\zeta,\omega^2)=\frac{6}{2\sqrt{2N_c}}x(1-x)\sum_{n=0}^\infty\sum_{l=0}^{n+1}
B_{nl}(\omega^2)C_n^{3/2}(2x-1)P_l(2\zeta-1),
\end{eqnarray}
where $B_{nl}(\omega^2)$ are the $\omega^2$-dependent coefficients, $N_c=3$ is the number of colors, and
$l=0,1, 2$,... denote the $S$-wave , $P$-wave, $D$-wave,... components, respectively.

The time-like form factor $B_{0l}(\omega^2)$, which normalizes each of the partial-wave component,
contains both resonant and nonresonant contributions. Some form factors, such as the time-like
pion form factor that receives contributions from the series of $\rho$ resonances, have been
constrained stringently by experimental data~\cite{prd86-032013}. The other coefficients
$B_{nl}(\omega^2)$, referred to as the Gegenbauer moments, are still quite uncertain
due to a lack of systematic nonperturbative studies. Note that these Gegenbauer moments differ
from those in the DA for a specific resonance which strongly decays into
the meson pair, because, as stated above, a two-meson
DA collects contributions from a series of resonances as well as nonresonant contributions.
Moreover, they are $\omega^2$-dependent, a feature dramatically distinct from the
Gegenbauer moments for a meson DA. It has been observed~\cite{plb763-29} that
the Gegenbauer moments of a $P$-wave di-pion DA differ from those of the $\rho(770)$ meson DA.
Therefore, it is essential to determine the Gegenbauer moments for two-meson DAs in order to
improve the precision of theoretical predictions for multi-body $B$ meson decays
in factorization frameworks.

We will perform a global fit of the Gegenbauer moments in two-meson DAs to measured
branching ratios and direct $CP$ asymmetries in three-body charmless hadronic $B$ meson decays
$B\to VP_3\to P_1P_2P_3$ in the PQCD approach, where $V$ stands for an intermediate vector resonance,
and $P_i$, $i=1,2,3$, stand for final-state pseudoscalar mesons. As the first attempt to a global
determination of two-meson DAs, we focus on the $P$-wave components, and employ the LO PQCD
factorization formulas for decay amplitudes.
We establish a Gegenbauer-moment-independent database, by means of which each decay amplitude
is expressed as a combination of the relevant Gegenbauer moments in the two-meson DAs. The Gegenbauer
moments in the DAs for the mesons $P_3=\pi,K$ are input from the global analysis of two-body $B$
meson decays in Ref.~\cite{2012-15074}. The leading-twist (twist-2) and next-to-leading-twist
(twist-3) DAs for the pairs $P_1P_2=\pi\pi, K\pi$ and $KK$ with the intermediate vector mesons
$V=\rho, K^*$ and $\phi$, respectively, are then fixed in the global fit. Because
the current data for three-body $B$ meson decays are not yet precise enough to determine
the $\omega^2$ dependence of the Gegenbauer moments, we first treat them as constant parameters
defined at the initial scale 1 GeV. One or two Gegenbauer moments for each of the above
two-meson DAs are obtained with satisfactory fit quality, depending on the abundance of available
data. It is noticed that the results and the precision of the extracted
two-meson DAs depend on the number of the Gegenbauer moments considered in the fit:
when more Gegenbauer moments are introduced into the $K\pi$ DAs, the quality of the fit is improved
at the cost of amplified uncertainties for fit outcomes.

The determined Gegenbauer
moments are then employed to make predictions for those observables, whose data are excluded in the
fit due to larger experimental errors. A general consistency between our predictions and
data for various modes is achieved, except those which suffer significant subleading corrections
according to previous PQCD studies,
such as the $B^0 \to \pi^0(\rho^0\to)\pi \pi$ decay~\cite{prd74-094020,Epjc72-1923}.
The consistency hints the validity of the PQCD formalism for
three-body hadronic $B$ meson decays and the universality of the nonperturbative two-meson DAs.
The $\pi\pi, K\pi$ and $KK$ twist-2 and twist-3 DAs presented in this work are
ready for applications to PQCD investigations of other multi-body $B$ meson decays involving the same
meson pairs. Our formalism can be extended to global fits for other two-meson DAs of various
partial waves straightforwardly. It can be also generalized to include
higher-order and/or higher-power corrections to PQCD factorization formulas~\cite{Li:2012nk},
when they are available, so that more accurate two-meson DAs are attainable in a systematic way.

As a more ambitious attempt, we explore the dependence of the Gegenbauer moments in
the di-pion DAs on the pion-pair invariant mass squared $\omega^2$.
It is unlikely to determine the exact $\omega^2$ dependence
from current data, so we simply parametrize the Gegenbauer moments up to the first power in $\omega^2$,
following their series expansion derived in Ref.~\cite{MP}. The global fit shows
that at least the linear term in one of the twist-3 di-pion DAs can be constrained
effectively. It indicates that the determination of the $\omega^2$-dependent Gegenbauer moments
in two-meson DAs is promising, when data become more precise in the future.

The rest of the paper is organized as follows. The kinematic variables for three-body hadronic
$B$ meson decays are defined in Sec.~II, where the dependence on final-state meson masses is
included to describe the phase space accurately. The parton kinematics and hard kernels are then refined,
such that $SU(3)$ symmetry breaking effects in the decays can be evaluated more precisely. The considered
two-meson $P$-wave DAs are parametrized, whose normalization form factors are assumed to take
the relativistic Breit-Wigner (RBW) model~\cite{epjc78-1019} or the Gounaris-Sakurai (GS)
model~\cite{prl21-244}. We explain how to perform the global fit, present and discuss the
numerical results, and try to extract the $\omega^2$ dependence of the Gegenbauer moments in
Sec.~III, which is followed by the Conclusion. We collect the PQCD factorization
formulas for the decay amplitudes with numerous refined hard kernels in the Appendix.

\section{FRAMEWORK}\label{sec:2}
\subsection{Kinematics}

Consider the charmless $B$ meson decay into three pseudoscalar mesons via
a vector intermediate resonance,
$B(p_B)\rightarrow V(p)P_3(p_3)\rightarrow P_1(p_1)P_2(p_2)P_3(p_3)$,
with the meson momenta $p_B=p+p_3$ and $p=p_1+p_2$.
We work in the $B$ meson rest frame and parametrize the relevant momenta in the light-cone coordinates as
\begin{eqnarray}
p_{B}&=&\frac{m_{B}}{\sqrt 2}(1,1,\textbf{0}_{\rm T}), ~~~~~~\quad k_{B}=\left(0,x_B \frac{m_{B}}{\sqrt2} ,\textbf{k}_{B \rm T}\right),\nonumber\\
p&=&\frac{m_{B}}{\sqrt2}(f_{+},f_{-},\textbf{0}_{\rm T}), ~\quad k= \left( z f_{+}\frac{m_{B}}{\sqrt2},0,\textbf{k}_{\rm T}\right),\nonumber\\
p_3&=&\frac{m_{B}}{\sqrt 2}(g_{-},g_{+},\textbf{0}_{\rm T}), ~\quad k_3=\left(0,x_3 g_{+} \frac{m_B}{\sqrt{2}},\textbf{k}_{3{\rm T}}\right),\label{mom-B-k}
\end{eqnarray}
where $m_{B}$ is the $B$ meson mass, and $k_{B}, k$ and $k_3$ are the valence quark momenta in the $B$ meson,
the meson pair, and the bachelor meson with the parton momentum fraction (transverse momenta)
$x_B, z$ and $x_3$ (${k}_{B \rm T}, {k}_{\rm T}$ and ${k}_{3{\rm T}}$), respectively. That is, we have
chosen the frame, such that the meson pair and the bachelor meson move in the directions $n=(1,0,0_{\rm T})$
and $v=(0,1,0_{\rm T})$, respectively. Since the parton momentum $k$ ($k_3$) is aligned with the meson pair
(bachelor meson), its small minus (plus) component has been neglected. We have also dropped the plus component $k_B^+$,
because it does not appear in the hard kernels for dominant factorizable contributions.
In the above expressions, the functions $f_{\pm}$ and $g_{\pm}$ are written as
\begin{eqnarray}
f_{\pm}&=&\frac{1}{2}\left(1+\eta-r_3\pm\sqrt{(1-\eta)^2-2r_3(1+\eta)+r_3^2}\right),\nonumber\\
g_{\pm}&=&\frac{1}{2}\left(1-\eta+r_3\pm\sqrt{(1-\eta)^2-2r_3(1+\eta)+r_3^2}\right),\label{fg}
\end{eqnarray}
with the ratio $r_3=m_{P_3}^2/m^2_{B_{(s)}}$ and $\eta=\omega^2/m^2_{B_{(s)}}$, $m_{P_3}$ being the bachelor meson mass
and  $\omega^2=p^2$ being the invariant mass squared of the meson pair.
For a $P$-wave meson pair, we introduce the longitudinal polarization vector
\begin{eqnarray}\label{eq:pq1}
\epsilon=\frac{1}{\sqrt{2\eta}}(f_{+},-f_{-},\textbf{0}_{\rm T}).
\end{eqnarray}

We derive the meson momenta $p_1$ and $p_2$,
\begin{eqnarray}\label{eq:p1p2}
 p_1&=&\left((\zeta+\frac{r_1-r_2}{2\eta})f_{+}\frac{m_B}{\sqrt{2}},
 (1-\zeta+\frac{r_1-r_2}{2\eta})f_{-}\frac{m_B}{\sqrt{2}}, \textbf{p}_{\rm T}\right), \nonumber\\
 p_2&=&\left((1-\zeta-\frac{r_1-r_2}{2\eta})f_{+}\frac{m_B}{\sqrt{2}},
 (\zeta-\frac{r_1-r_2}{2\eta})f_{-}\frac{m_B}{\sqrt{2}}, -\textbf{p}_{\rm T}\right),\nonumber\\
 p_{\rm T}^2&=&\zeta(1-\zeta)\omega^2+\frac{(m_{P_1}^2-m_{P_2}^2)^2}{4\omega^2}-\frac{m^2_{P_1}+m^2_{P_2}}{2},
\end{eqnarray}
from the relation $p=p_1+p_2$ and the on-shell conditions $p_i^{2}=m_{P_i}^{2}$, $i=1,2$,
with the mass ratios $r_{1,2}=m_{P_1,P_2}^2/m^2_B$. The variable
$\zeta+(r_1-r_2)/(2\eta)=p_1^+/p^+$ bears the meaning of the meson momentum fraction up to corrections
from the final-state meson masses.
Alternatively, one can define the polar angle $\theta$ of the meson $P_{1}$ in the $P_1P_2$ pair rest frame.
The transformation between the $B$ meson rest frame
and the meson pair rest frame leads to the relation between the meson momentum fraction $\zeta$ and
the polar angle $\theta$,
%One can obtain the relation between $\zeta$ and the polar angle $\theta$ in the dimeson rest frame,
\begin{eqnarray}\label{eq:cos}
2\zeta-1=\sqrt{1-2\frac{r_1+r_2}{\eta}+\frac{(r_1-r_2)^2}{\eta^2}}\cos\theta,
\end{eqnarray}
with the bounds
\begin{eqnarray}
\zeta_{\text{max,min}}=\frac{1}{2}\left[1\pm\sqrt{1-2\frac{r_1+r_2}{\eta}+\frac{(r_1-r_2)^2}{\eta^2}}\right].
\end{eqnarray}
We emphasize that the parametrization with the exact dependence on the final-state meson masses in
Eq.~(\ref{eq:p1p2}) is crucial for establishing Eq.~(\ref{eq:cos}), such that the Legendre
polynomials in Eq.~(\ref{2mda}) correspond to the partial waves of the meson pair exactly.

\begin{figure}[tbp]
%\vspace{-1cm}
\centerline{\epsfxsize=14cm \epsffile{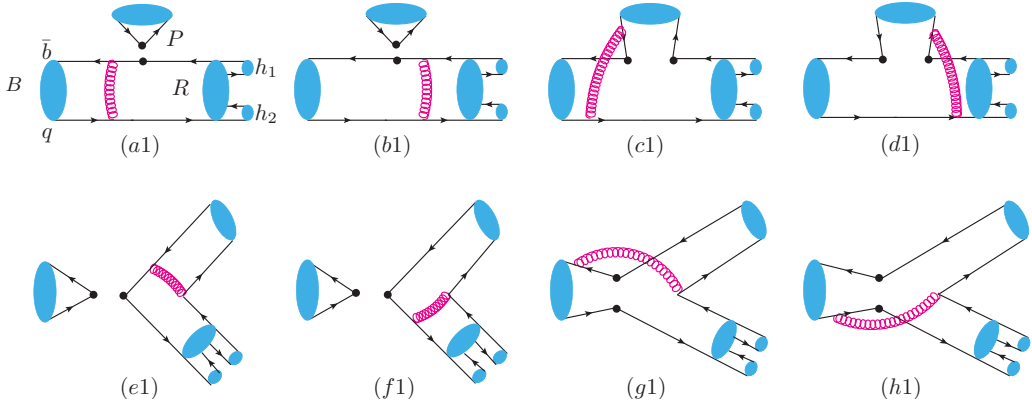}}
%\vspace{0.3cm}
%\vspace{-6cm}
\caption{LO diagrams for the three-body decays $B \to V P_3\to P_1P_2P_3$
with the light quarks $q=u,d,s$, where the symbol $\bullet$ represents the weak vertex.}
\label{fig:fig1}
\end{figure}

\begin{figure}[tbp]
%\vspace{-1cm}
\centerline{\epsfxsize=14cm \epsffile{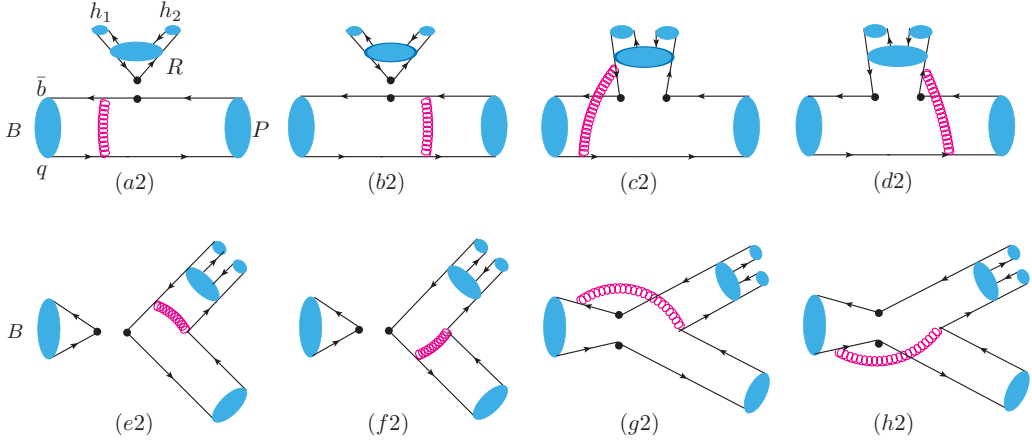}}
%\vspace{0.3cm}
%\vspace{-6cm}
\caption{More LO diagrams for the three-body decays $B \to V P_3\to P_1P_2P_3$.}
\label{fig:fig2}
\end{figure}

The branching ratio for a three-body $B$ meson decay is given by~\cite{pdg2020}
\begin{eqnarray}\label{eq:br}
\int d\mathcal{B}=\frac{\tau_B m_B}{256\pi^3} \int^1_{(\sqrt{r_1}+\sqrt{r_2})^2} d\eta \sqrt{(1-\eta)^2-2r_3(1+\eta)+r^2_3}\int^{\zeta_{\text{max}}}_{\zeta_{\text{min}}}d\zeta|\mathcal{A}|^2,
\end{eqnarray}
with the $B$ meson lifetime $\tau_B$.
The direct $CP$ asymmetry $A_{CP}$ is then defined as
\begin{eqnarray}
A_{CP}&=&\frac{{\cal B}(\bar B\to \bar f)-{\cal B}( B\to  f)}{{\cal B}(\bar B\to \bar f)+{\cal B}( B\to  f)}.
%\non
%&=& \frac{|{\cal A}(\bar B\to \bar f)|^2-|{\cal A}( B\to  f)|^2}{|{\cal A}(\bar B\to \bar f)|^2+|{\cal A}( B\to  f)|^2}.
\end{eqnarray}
The decay amplitude $\mathcal{A}$, according to the factorization
theorem stated in the Introduction, is expressed as
\begin{eqnarray}
\mathcal{A}= \Phi_B \otimes H\otimes \Phi_{P_1P_2} \otimes \Phi_{P_3},
\end{eqnarray}
where $\Phi_B$ ($\Phi_{P_3}$) is the $B$ (bachelor) meson DA, and the two-meson DA $\Phi_{P_1P_2}$
absorbs the nonperturbative dynamics in the production of the meson pair $P_1P_2$.
The symbol $\otimes$ denotes the convolution of the above factors in parton momenta.
The LO diagrams for the hard kernel $H$
are displayed in Figs.~\ref{fig:fig1} and \ref{fig:fig2}, where Figs.~\ref{fig:fig1}(a)-(d)
(Figs.~\ref{fig:fig2}(a)-(d)) are associated with the $B\to P_1P_2$ ($B\to P_3$) transition,
and Figs.~\ref{fig:fig1}(e)-(h) and Figs.~\ref{fig:fig2}(e)-(h) with the annihilation contributions.

\subsection{Distribution Amplitudes}\label{sec:22}

The light-cone hadronic matrix element for a $B$ meson is parametrized
as~\cite{prd63-054008,prd65-014007,epjc28-515,ppnp51-85,Prd85-094003}
\begin{eqnarray}
\int d^4z e^{i{\bf k_1}\cdot z}\langle0|q_{\beta}(z)\bar{b}_{\alpha}(0)|B(p_{B})\rangle=
\frac{i}{\sqrt{2N_c}}\left\{({p \hspace{-2.0truemm}/}_{B}+m_{B})\gamma_{5}\left[\phi_{B}({\bf k_1})-
\frac{{n \hspace{-2.0truemm}/}-{v \hspace{-2.0truemm}/}}{\sqrt{2}}\bar{\phi}_{B}({\bf k_1})\right]\right\}
_{\beta\alpha},
\end{eqnarray}
where $q$ represents a $u$, $d$ or $s$ quark. The two wave functions $\phi_{B}$ and
$\bar \phi_{B}$ in the above decomposition, related to $\phi_{B}^+$ and $\phi_{B}^-$ defined in the
literature~\cite{GN} via $\phi_{B}=(\phi_{B}^++\phi_{B}^-)/2$ and $\bar \phi_{B}=(\phi_{B}^+-\phi_{B}^-)/2$,
obey the normalization conditions
\begin{eqnarray}
\int \frac{d^4{\bf k_1}}{(2\pi)^4}\phi_{B}({\bf k_1})=\frac{f_{B}}{2\sqrt{2N_c}},\;\;\;\;
\int\frac{d^4{\bf k_1}}{(2\pi)^4}\bar{\phi}_{B}({\bf k_1})=0.
\end{eqnarray}
It has been shown that the contribution from $\bar \phi_{B}$ is of next-to-leading power
and numerically suppressed~\cite{prd65-014007,epjc28-515,Prd103-056006}, compared to the leading-power contribution from $\phi_B$.
Taking the PQCD evaluation of the  $B\to\pi$ transition form factor $F_0^{B\to \pi}$ in Ref.~\cite{Prd103-056006}
as an example, we find that
%\begin{eqnarray}
%F_0^{B\to \pi}(0)=
%\left\{\begin{array}{ll}
%0.190\pm0.006(\omega_B)\pm0.008(m_0^{\pi})\pm0.008(a_2^{\pi}),   & \phi_B,\\
%0.041\pm0.001(\omega_B)\pm0.000(m_0^{\pi})\pm0.005(a_2^{\pi}),   & \bar{\phi}_{B}.\\%\cite{PRD82-034026}, \\
%\end{array} \right.
%\end{eqnarray}
%Namely,
the  $\bar{\phi}_{B}$ contribution to $F_0^{B\to \pi}$ is about 20\% of the $\phi_B$ one.
The higher-twist $B$ meson DAs have been
systematically investigated in the heavy quark effective theory~\cite{jhep05-022},
which are decomposed according to definite twist and conformal spin assignments up to twist 6.
In principle, all the next-to-leading-power sources should be included
for a consistent and complete analysis, which, however, goes beyond the scope of the present formalism.
Therefore, we focus only on the leading-power component
\begin{eqnarray}
\Phi_B= \frac{i}{\sqrt{2N_c}} ({ p \hspace{-2.0truemm}/ }_B +m_B) \gamma_5 \phi_B ( x,b), \label{bmeson}
\end{eqnarray}
with the impact parameter $b$ being  conjugate to the parton transverse momentum $k_{B \rm T}$.
The $B$ meson DA $\phi_B (x,b)$ is chosen as the model form widely adopted in
the PQCD approach~\cite{prd63-054008,prd65-014007,epjc28-515,ppnp51-85,Prd85-094003,Li:2012md},
\begin{eqnarray}
\phi_B(x,b)&=& N_B x^2(1-x)^2\mathrm{exp} \left  [ -\frac{m_B^2 x^2}{2 \omega_{B}^2} -\frac{1}{2} (\omega_{B} b)^2\right] ,
 \label{phib}
\end{eqnarray}
where the constant $N_B$ is related to the $B$ meson decay constant $f_B$
through the normalization condition $\int_0^1dx \; \phi_B(x,b=0)=f_B/(2\sqrt{2N_c})$.
The shape parameter takes the values $\omega_B = 0.40$ GeV for $B^+,B^0$ mesons and $\omega_{B_s}=0.48$
GeV~\cite{prd63-054008,plb504-6,prd63-074009,2012-15074} for a $B^0_s$ meson with 10\% variation in the
numerical study below.

The light-cone matrix elements for a pseudoscalar meson is decomposed, up to twist 3, into~\cite{prd65-014007,epjc28-515}
\begin{eqnarray}
\Phi_{P}\equiv \frac{i}{\sqrt{2N_C}}\gamma_5
                    \left [{ p \hspace{-2.0truemm}/ }_3 \phi_{P}^{A}(x_3)+m_{03} \phi_{P}^{P}(x_3)
                    + m_{03} ({ n \hspace{-2.2truemm}/ }
                    { v \hspace{-2.2truemm}/ } - 1)\phi_{P}^{T}(x_3)\right ],
\end{eqnarray}
with $P=\pi, K$ and the chiral scale $m_{03}$. The pion and kaon DAs
have been determined at the scale 1 GeV in a recent global analysis~\cite{2012-15074}
based on LO PQCD factorization formulas, which is at the same level of accuracy as this work.
The results are quoted as
\begin{eqnarray}
 \phi_{\pi}^A(x) &=& \frac{3f_{\pi}}{\sqrt{6}} x(1-x)[ 1 +a_2^{\pi}C_2^{3/2}(2x-1)+a_4^{\pi}C_4^{3/2}(2x-1)], \nonumber\\
 \phi_{\pi}^P(x) &=& \frac{f_{\pi}}{2\sqrt{6}}[1 +a_{2P}^{\pi}C_2^{1/2}(2x-1)], \nonumber\\
 \phi_{\pi}^T(x) &=& \frac{f_{\pi}}{2\sqrt{6}}(1-2x)[1+a_{2T}^{\pi}(10x^2-10x+1)],\nonumber\\
 \phi_{K}^A(x) &=& \frac{3f_{K}}{\sqrt{6}}x(1-x)[1+a_1^{K}C_1^{3/2}(2x-1)+a_2^{K}C_2^{3/2}(2x-1)+a_4^{K}C_4^{3/2}(2x-1)],\nonumber \\
 \phi_{K}^P(x) &=& \frac{f_{K}}{2\sqrt{6}} [1+a_{2P}^{K}C_2^{1/2}(2x-1)],\nonumber \\
 \phi_{K}^T(x) &=& -\frac{f_{K}}{2\sqrt{6}}[C_1^{1/2}(x)+a_{2T}^{K}C_3^{1/2}(x)],
 \label{eq:pkda}
\end{eqnarray}
where the Gegenbauer polynomials are defined as
\begin{eqnarray}
C_1^{3/2}(t)=3t, ~\quad\quad C_2^{3/2}(t)=\frac{3}{2}(5t^2-1), ~\quad\quad C_4^{3/2}(t)=\frac{15}{8}(1-14t^2+21t^4),
\end{eqnarray}
and the Gegenbauer moments in Eq.~(\ref{eq:pkda}) are summarized as follows,
\begin{eqnarray}
a_2^{\pi}&=&0.64\pm0.08,\quad a^{\pi}_{4}=-0.41\pm0.10,\quad a^{\pi}_{2P}=1.08\pm0.15,\quad a^{\pi}_{2T}=-0.48\pm0.33,\non
a^{K}_{1}&=&0.33\pm0.08,\quad a^{K}_{2}=0.28\pm0.10, \quad a^{K}_{4}=-0.40\pm0.07,\quad a^{K}_{2P}=0.24, \quad a^{K}_{2T}=0.35.
\label{eq:genpik}
\end{eqnarray}
Note that the twist-3 DAs $\phi_K^P$ and $\phi_K^T$, which were not obtained in Ref.~\cite{2012-15074},
come from sum-rule calculations~\cite{prd76-074018}.

As stated before, we focus on the $P$-wave components in Eq.~(\ref{2mda}) proportional to $P_{l=1}(2\zeta-1)=2\zeta-1$.
The corresponding light-cone matrix element for a longitudinal meson pair is decomposed, up to the twist 3, into~\cite{plb763-29}
\begin{eqnarray}
\Phi_{P_1P_2}(x,\zeta,\omega^2)=\frac{1}{\sqrt{2N_c}}\left[\omega{ \epsilon \hspace{-1.5truemm}/ }
\phi_{P_1P_2}^0(x,\omega^2)+\omega\phi_{P_1P_2}^{s}(x,\omega^2)
+\frac{{p\hspace{-1.5truemm}/}_1{p\hspace{-1.5truemm}/}_2
  -{p\hspace{-1.5truemm}/}_2{p\hspace{-1.5truemm}/}_1}{\omega(2\zeta-1)}\phi_{P_1P_2}^t(x,\omega^2)
\right](2\zeta-1),
\label{eq:phifunc}
\end{eqnarray}
where the two-meson DAs for the $\pi\pi, KK$ and $K\pi$ pairs are parametrized as
\begin{eqnarray}
\phi_{\pi\pi}^0(x,\omega^2)&=&\frac{3F_{\pi\pi}^{\parallel}(\omega^2)}{\sqrt{2N_c}}x(1-x)\left[1
+a^0_{2\rho}C_2^{3/2}(2x-1)\right]\;,\nonumber\\ %\label{eq:pild0}
\phi_{\pi\pi}^{s}(x,\omega^2)&=&\frac{3F_{\pi\pi}^{\perp}(\omega^2)}{2\sqrt{2N_c}}(1-2x)\left[1
+a^s_{2\rho}(10x^2-10x+1)\right]\;,\nonumber\\%\label{eq:pilds}
\phi_{\pi\pi}^t(x,\omega^2)&=&\frac{3F_{\pi\pi}^{\perp}(\omega^2)}{2\sqrt{2N_c}}(1-2x)^2\left[1
+a^t_{2\rho}C_2^{3/2}(2x-1)\right]\;,\nonumber\\ %\label{eq:pildt}
\phi_{K\pi}^0(x,\omega^2)&=&\frac{3F_{K\pi}^{\parallel}(\omega^2)}{\sqrt{2N_c}} x(1-x)
\left[1+a_{1K^*}^{0}C_1^{3/2}(2x-1)+a_{2K^*}^{0}C_2^{3/2}(2x-1)+a_{4K^*}^{0}C_4^{3/2}(2x-1)\right]\;,\nonumber\\%\label{eq:pikld0}
\phi_{K\pi}^s(x,\omega^2)&=&\frac{3F_{K\pi}^{\perp}(\omega^2)}{2\sqrt{2N_c}}(1-2x)\;,\nonumber\\%\label{eq:piklds}
\phi_{K\pi}^t(x,\omega^2)&=&\frac{3F_{K\pi}^{\perp}(\omega^2)}{2\sqrt{2N_c}}(1-2x)^2\;,\nonumber\\
\phi_{KK}^0(x,\omega^2)&=&\frac{3F_{KK}^{\parallel}(\omega^2)}{\sqrt{2N_c}}x(1-x)\left[1
+a^0_{2\phi}C_2^{3/2}(2x-1)\right]\;,\nonumber\\ %\label{eq:kkld0}
\phi_{KK}^{s}(x,\omega^2)&=&\frac{3F_{KK}^{\perp}(\omega^2)}{2\sqrt{2N_c}}(1-2x)\;,\nonumber\\ %\label{eq:kklds}
\phi_{KK}^t(x,\omega^2)&=&\frac{3F_{KK}^{\perp}(\omega^2)}{2\sqrt{2N_c}}(1-2x)^2\;. %\label{eq:kkldt}
\label{eq:pikldt}
\end{eqnarray}
The Gegenbauer moments $a^{0,s,t}_{2\rho}$, $a_{1K^*,2K^*,4K^*}^{0}$,
and $a^0_{2\phi}$ will be determined in a global analysis in the
next section. Since the current data are not yet precise enough for fixing the Gegenbauer moments in
the twist-3 DAs $\phi_{K\pi}^{s,t}$ and $\phi_{KK}^{s,t}$, they have been
set to the asymptotic forms.

%\begin{table}
%\centering%% Table 1
%\caption{Input parameters for the considered intermediate states.}
%\label{Tab:pa}
%\setlength{\tabcolsep}{5mm}{
%\begin{tabular}{llllll} \hline
%{\rm Resonances} &Masses~(MeV) &Widths~(MeV) &$J^{P}$&Models&Sources\\ \hline
%$\rho(770)$ &$775.02\pm0.31$ &$149.59\pm0.67$ &$1^{-}$&GS &BABAR~\cite{prd86-032013}\\
%$K^*(892)$ &$895.55\pm0.20$&$47.3\pm0.5$& $1^{-}$ & RBW &LHCb~\cite{epjc78-1019}\\
%$\phi(1020)$ &$1019.461\pm0.016$&$4.249\pm0.013$& $1^{-}$ & RBW &PDG \cite{pdg2020}\\
%\hline
%\end{tabular}}
%\end{table}

The elastic rescattering effects in a final-state meson pair can be absorbed into the
time-like form factors $F^{\parallel,\perp}(\omega^2)$, namely, the leading Gegenbauer moment $B_{00}(\omega^2)$
in a two-meson DA according to the Watson theorem~\cite{pr88-1163}.
The resonant contribution from a $\rho$ meson with a broad width is usually
parameterized as the GS model~\cite{prl21-244} based on the Breit-Wigner (BW)
function~\cite{BW-model} in experimental investigations of
three-body hadronic $B$ meson decays, which interprets observed structures beyond the $\rho(770)$
resonance in terms of heavier isovector vector mesons.
Taking the $\rho$-$\omega$ interference and excited state contributions into account, we have
the form factor~\cite{prd86-032013,prd95056008,plb763-29}
\begin{eqnarray}
F^\parallel_{\pi\pi}(\omega^2)= \left [ {\rm GS}_\rho(\omega^2,m_{\rho},\Gamma_{\rho})
\frac{1+c_{\omega} {\rm BW}_{\omega}(\omega^2,m_{\omega},\Gamma_{\omega})}{1+c_{\omega}}
+\Sigma c_j {\rm GS}_j(\omega^2,m_j,\Gamma_j)\right] \left( 1+\Sigma c_j\right)^{-1},
\label{GS}
\end{eqnarray}
where $m_{\rho,\omega,j}$ ($\Gamma_{\rho,\omega,j}$),
$j=\rho^{\prime}(1450),\rho^{\prime \prime}(1700)$ and $\rho^{\prime \prime \prime}(2254)$,
are the masses (decay widths) of the series of resonances, and $c_{\omega,j}$ are the
weights associated with the corresponding resonances.
The function ${\rm GS}_\rho(s,m_{\rho},\Gamma_{\rho})$ is given by
\begin{equation}
{\rm GS}_\rho(\omega^2, m_\rho, \Gamma_\rho) =
\frac{m_\rho^2 [ 1 + d(m_\rho) \Gamma_\rho/m_\rho ] }{m_\rho^2 - \omega^2 + f(\omega^2, m_\rho, \Gamma_\rho)
- i m_\rho \Gamma (\omega^2, m_\rho, \Gamma_\rho)},
\end{equation}
with the factors
\begin{eqnarray}
d(m_\rho) &=& \frac{3}{\pi} \frac{m_\pi^2}{k^2(m^2_\rho)} \ln \left( \frac{m_\rho+2 k(m^2_\rho)}{2 m_\pi} \right)
   + \frac{m_\rho}{2\pi  k(m_\rho^2)}
   - \frac{m_\pi^2  m_\rho}{\pi k^3(m^2_\rho)},\non
f(\omega^2, m_\rho, \Gamma_\rho) &=& \frac{\Gamma_\rho  m^2_\rho}{k^3(m^2_\rho)} \left\{ k^2(\omega^2) [ h(\omega^2)-h(m^2_\rho) ]
+ (m^2_\rho-\omega^2) k^2(m^2_\rho)  h'(m^2_\rho)\right\},\non
\Gamma (\omega^2, m_\rho, \Gamma_\rho) &=& \Gamma_\rho  \frac{\omega^2}{m^2_\rho}
\left[ \frac{\beta_\pi (\omega^2) }{ \beta_\pi (m^2_\rho) } \right]^3,
\end{eqnarray}
where the functions $k(\omega^2)$, $h(\omega^2)$ and $\beta_\pi (\omega^2)$ are expressed as
\begin{eqnarray}
k(\omega^2) &=& \frac{1}{2} \sqrt{\omega^2}  \beta_\pi (\omega^2),\non
h(\omega^2) &=& \frac{2}{\pi}  \frac{k(\omega^2)}{\sqrt{\omega^2}}
\ln \left( \frac{\sqrt{\omega^2}+2 k(\omega^2)}{2 m_\pi} \right),\non
\beta_\pi (\omega^2) &=& \sqrt{1 - \frac{4m_\pi^2}{\omega^2}}.
\end{eqnarray}
The function  ${\rm BW}_{\omega}(\omega^2,m_{\omega},\Gamma_{\omega})$ for the $\omega$ resonance
takes the standard BW form~\cite{BW-model}.

We apply the RBW line shape for contributions from the intermediate resonances
$K^*$ and $\phi$ of narrow widths to the form factors~\cite{epjc78-1019,epjc80517,epjc8191,epjc7937},
\begin{eqnarray}
\label{BRW}
F^\parallel_{K\pi,KK}(\omega^2)&=&\frac{m_{K^*,\phi}^2}{m^2_{K^*,\phi} -\omega^2-im_{K^*,\phi}\Gamma_{K^*,\phi}(\omega^2)},
\end{eqnarray}
with the mass-dependent widths
\begin{eqnarray}
\label{BRWl}
\Gamma_{K^*,\phi}(\omega^2)&=&\Gamma_{K^*,\phi}\left(\frac{m_{K^*,\phi}}{\omega}\right)
\left(\frac{|\vec{p}_1|}{|\vec{p}_0|}\right)^{(2L_R+1)},
\end{eqnarray}
%$i=K^*$ and $\phi$.
where the masses $m_{K^*,\phi}$ and the widths $\Gamma_{K^*,\phi}$ of the $K^*$ and $\phi$
resonances, respectively, take the values in~\cite{pdg2020}. The magnitude of the spatial momentum of the meson $P_1$,
\begin{eqnarray}
|\vec{p}_1|=\frac{\sqrt{\lambda(\omega^2,m_{P_1}^2,m_{P_2}^2)}}{2\omega},
\end{eqnarray}
with the K$\ddot{a}$ll$\acute{e}$n function $\lambda (a,b,c)= a^2+b^2+c^2-2(ab+ac+bc)$,
is measured in the rest frame of the resonance, and $|\vec{p}_0|$ is its value at the resonance mass.
The orbital angular momentum $L_R$ in the two-meson system is set to $L_R=1$ for a $P$-wave state.
Due to the limited knowledge on the form factors $F^{\perp}(\omega^2)$, we assume the ratio
$F^{\perp}_i(\omega^2)/F^{\parallel}_i(\omega^2)
\approx (f^T_i/f_i)$~\cite{plb763-29}, $i=\rho, K^*$ and $\phi$,
with $f_i^T$ ($f_i$) being the tensor (vector) decay constants of the intermediate resonances.

\section{Numerical Analysis}

\subsection{Global Fit}
We specify the parameters adopted in the numerical analysis below, including the masses (in units of GeV)~\cite{pdg2020}
\begin{eqnarray}
m_{B}&=&5.280, \quad m_{B_s}=5.367, \quad m_b=4.8, \quad ~~~~m_{K^\pm}=0.494,\nonumber\\
 m_{K^0}&=&0.498, \quad m_{\pi^{\pm}}=0.140, \quad m_{\pi^0}=0.135,
\end{eqnarray}
and the decay constants (in units of GeV) and the $B$ meson lifetimes (in units of ps)~\cite{prd76-074018,prd95056008}
\begin{eqnarray}
f_B&=&0.21, ~~~\quad f_{B_s}=0.23, ~~~~~~~~~\quad f_{\rho}=0.216 , ~~~~\quad f^T_{\rho}=0.184,\nonumber\\
f_{\phi(1020)}&=&0.215, ~\quad f_{\phi(1020)}^T=0.186, \quad f_{K^*}=0.217, ~\quad f^T_{K^*}=0.185,\nonumber\\
\tau_{B^0}&=&1.519,~\quad \tau_{B^{\pm}}=1.638, ~~~~~~\quad \tau_{B_{s}}=1.512.
\end{eqnarray}
The Wolfenstein parameters in the Cabibbo-Kobayashi-Maskawa (CKM) matrix take the values in Ref.~\cite{pdg2018}:
$A=0.836\pm0.015, \lambda=0.22453\pm 0.00044$, $\bar{\rho} = 0.122^{+0.018}_{-0.017}$ and $\bar{\eta}= 0.355^{+0.012}_{-0.011}$.

We stress that $\omega_{B_{(s)}}$ in the $B_{(s)}$ meson DA, as an overall parameter,
cannot be determined in our global analysis, but must be treated as an input. This is why we take its
value extracted from the $B_{(s)}$ meson transition form factors in lattice QCD and light-cone sum rules,
which has been also verified by the global study of two-body charmless hadronic $B$ meson decays in~\cite{2012-15074}.
The value of $\omega_{B_{(s)}}$, together with the corresponding pion and kaon DAs fixed in~\cite{2012-15074}, are
then input into the present work on the three-body $B$ meson decays for consistency. If the shape parameter
$\omega_{B_{(s)}}$ is changed, the pion and kaon DAs need to be refitted accordingly, before they can be
employed to constrain the two-meson DAs. Fortunately, the variation of
$\omega_{B_{(s)}}$ causes less than $30\%$ uncertainties for most of the branching ratios and negligible effects
on the direct $CP$ asymmetries as seen later, and is thus
not expected to make a significant impact on the determination of the two-meson DAs.
Hence, we focus only on the uncertainties of the Gegenbauer moments in the two-meson DAs
propagated from experimental data here.

Equation~(\ref{eq:pikldt}) suggests that the total amplitudes $\cal{A}$ for the
$B_{(s)} \to P(\pi \pi,\pi K, KK)$ decays with $P=\pi, K$
can be expanded in terms of the Gegenbauer moments of the two-meson DAs.
As a result, we decompose the squared amplitudes
\begin{eqnarray}
|{\cal A}_{\pi\pi}|^2 &= & M_{0\rho}+a^{0}_{2\rho}M_{1\rho}+(a^{0}_{2\rho})^2M_{2\rho}+a^{s}_{2\rho}M_{3\rho}+(a^{s}_{2\rho})^2M_{4\rho}\non
           &+&      a^{t}_{2\rho}M_{5\rho}+(a^{t}_{2\rho})^2 M_{6\rho}+a^{0}_{2\rho}a^{s}_{2\rho}M_{7\rho}+
                 a^{0}_{2\rho}a^{t}_{2\rho}M_{8\rho}+a^{s}_{2\rho}a^{t}_{2\rho}M_{9\rho},\nonumber\\
|{\cal A}_{K\pi}|^2  &= & M_{0K^*}+(a^{0}_{1K^*})M_{1K^*}+(a^{0}_{1K^*})^2M_{2K^*}+a^{0}_{2K^*}M_{3K^*}\non
           &+&      (a^{0}_{2K^*})^2M_{4K^*}+a^{0}_{4K^*} M_{5K^*}+(a^{0}_{4K^*})^2M_{6K^*}\non
           &+&      a^{0}_{1K^*}a^{0}_{2K^*}M_{7K^*}+a^{0}_{1K^*}a^{0}_{4K^*} M_{8K^*}
           +a^{0}_{2K^*}a^{0}_{4K^*}M_{9K^*},\nonumber\\
|{\cal A}_{KK}|^2 &= & M_{0\phi}+a^0_{2\phi}M_{1\phi}+(a^0_{2\phi})^2 M_{2\phi},\label{a}
\end{eqnarray}
into the linear combinations of the Gegenbauer moments $a^{0,s,t}_{2\rho}$, $a^{0}_{1K^*,2K^*,4K^*}$ and $a^0_{2\phi}$,
and their products. We then compute the coefficients $M$, which involve only the Gegenbauer polynomials,
to establish the database for the global fit.

\begin{table}[!htbh]
\caption{Fitted Gegenbauer moments for the twist-2 and twist-3 two-meson DAs.}
\label{tab:gen}
\begin{center}
\setlength{\tabcolsep}{3mm}{
\begin{tabular}{lcccccccccccccc}
		\hline
		 &$a^0_{2\rho}$             &$a^s_{2\rho}$                &$a^t_{2\rho}$                &$a^0_{2\phi}$               &    \\ \hline
\ fit \  &$0.08\pm0.13$                     &$-0.23\pm0.24$            &$-0.35\pm0.06$                   &$-0.31\pm0.19$             & \\  \hline
&$a_{1K^*}^{0}(\text{Scenario I})$            &$a_{2K^*}^{0}(\text{Scenario I})$  &$a_{1K^*}^{0}(\text{Scenario II})$            &$a_{2K^*}^{0}(\text{Scenario II})$     &$a_{4K^*}^{0}(\text{Scenario II})$ \\ \hline
\ fit \     &$0.31\pm0.16$ & $1.19\pm0.10$ &$0.57\pm0.20$                                &$1.13\pm0.32$              &$-0.85\pm0.16$ \\  \hline

\end{tabular}}
\end{center}
\end{table}

Similar to the proposal in~\cite{2012-15074}, we determine the Gegenbauer moments of
the two-meson DAs by fitting the formulas in Eq.~(\ref{a}) with the
Gegenbauer-moment-independent database to the
measured branching ratios ${\cal B}$ and direct $CP$ asymmetries ${\cal A}_{CP}$ of the
$B_{(s)} \to P(\rho\to)\pi\pi$, $B_{(s)} \to P(K^*\to)K\pi$ and $B_{(s)} \to P(\phi\to)KK$ decays.
We adopt the standard nonlinear least-$\chi^2$ (lsq) method~\cite{Peter:2020}, in which
the $\chi^2$ function is defined for $n$ pieces of experimental data
$v_i\pm \delta v_i$ with the errors  $\delta v_i$ and the corresponding theoretical
values $v^{\rm{th}}_i$ as
\begin{eqnarray} \label{eq:chi}
	\chi^2= \sum_{i=1}^{n}  \Big(\frac {v_i - v^{\rm{th}}_i}{\delta v_i}\Big)^2.
\end{eqnarray}
The theoretical values $v^{\rm{th}}_i$ are the functions of the fitted Gegenbauer moments in
the two-meson DAs. The lsq fit attempts to find the smallest $\chi^2$
by adjusting the fitted parameters that bring the theoretical values closest
to the data. The data with errors are treated as of the Gaussian type automatically
in the fit packages, and the errors of the fitted parameters and of the theoretical values
$v^{\rm{th}}_i$ come from experimental uncertainties by error propagation.

To minimize statistical uncertainties, we should include maximal amount of
data in the fit. On the other hand, those measurements with significance lower than 3$\sigma$
do not impose stringent constraints, and need not be taken into account in principle.
The data of those modes, which are affected by subleading contributions
manifestly based on the previous PQCD studies~\cite{prd74-094020,Epjc72-1923}, are also excluded, even
though they may have higher precision. The $B^0 \to \pi^0(\rho^0\to)\pi\pi$ decay, dominated by
the color-suppressed tree amplitude that is expected to receive substantial higher-order corrections~\cite{Li:2005kt},
is a typical example.

\subsection{Results}

%The global fit yields the Gegenbauer moments in Table~\ref{tab:gen}, where two (three) Gegenbauer moments
%$a_{1K^*}^{0}$ and $a_{2K^*}^{0}$ ($a_{1K^*}^{0}$, $a_{2K^*}^{0}$ and $a_{4K^*}^{0}$) are considered
%for the twist-2 $K\pi$ DA in Scenario I (II).

The Gegenbauer moments $a^0_{2\rho}$, $a^s_{2\rho}$ and $a^t_{2\rho}$ for the twist-2 and twist-3
$\pi\pi$ DAs in Table~\ref{tab:gen} are obtained from the fit to eight pieces of $B\to P(\rho\to)\pi\pi$
data marked by ''$\dagger$" in Tables~\ref{krho} and \ref{pirho} with $\chi^2 / d.o.f.=2.6$, whose errors
mainly arise from experimental uncertainties. We point out that the measured
$B^+ \to \pi^+(\rho^0\to)\pi\pi$ branching ratio, imposing a strong constraint on the
Gegenbauer moment $a^s_{2\rho}$, is considered in our fit, but the corresponding $B^+ \to \pi^+\rho^0$ data
were excluded in the global analysis of two-body hadronic $B$ meson decays~\cite{2012-15074}. It is seen
that our Gegenbauer moments differ from the corresponding ones of the $\rho(770)$ meson
DAs derived in QCD sum rules~\cite{ball98} as mentioned before: the $\pi\pi$ DAs contain the $\rho$-$\omega$
mixing effect and the contributions from higher $\rho$ resonances with finite widths via Eq.~(\ref{GS}),
so they need not to be the same as the $\rho(770)$ meson DAs. Our Gegenbauer moments also differ from
$a^0_{2\rho}=0.25$, $a^s_{2\rho}=0.75$ and $a^t_{2\rho}=-0.60$ chosen in Ref.~\cite{plb763-29}
for two reasons at least. First, only the $B \to K(\rho \to)\pi\pi$ data were employed to constrain the
$\pi\pi$ DAs in~\cite{plb763-29}, while the additional $B \to \pi(\rho \to)\pi\pi$ data are included
in our global analysis. Second, some $B \to K(\rho \to)\pi\pi$ data have been updated in
this work.

A single Gegenbauer moment $a^0_{2\phi}$ is introduced into the $KK$ twist-2 DA, and the twist-3 ones
have been set to their asymptotic forms,
since only two pieces of data from the $B\to K(\phi \to)KK$ decays in Table~\ref{kk} meet the required precision.
The value of $a^0_{2\phi}$, determined with $\chi^2 / d.o.f.=0.35$,
is distinct from, but still consistent with that of the $\phi$ meson DA in QCD sum rules~\cite{ball98}
within theoretical errors. Note that our $a^0_{2\phi}$ deviates from the value
$-0.50\pm 0.10$ adopted in Ref.~\cite{epjc79792},
where $B_s$ meson decays into charmonia plus a kaon pair were investigated.
The deviation is understandable, because the choice of $a^0_{2\phi}$ depends on models
for the uncertain charmonium DAs, as relevant data were accommodated.

The $K\pi$ DAs are determined in a fit to six pieces of $B_{(s)}\to P(K^*\to)K\pi$ data in Tables~\ref{kkstar}
and \ref{pikstar}. We first work on Scenario I, in which the two Gegenbauer moments $a_{1K^*}^{0}$ and
$a_{2K^*}^{0}$ of the twist-2 two-meson DA are fitted with $\chi^2 / d.o.f.=1.5$, and observe that $a_{2K^*}^{0}$
is slightly larger than unity as shown in  Table~\ref{tab:gen}. A larger moment is not favored in view of the
convergence of the Gegenbauer expansion. Therefore, one more Gegenbauer moment $a_{4K^*}^{0}$ is
added in Scenario II, and a fit with $\chi^2 / d.o.f.=1.4$ is attained. The resultant $a_{2K^*}^{0}$
decreases a bit but with amplified uncertainty , and $a_{4K^*}^{0}$ is smaller than the unity.
The measured $B^0_s \to K^\pm(K^{*\mp} \to)K\pi$ and $B^0_s \to \KorKbar^0 (\KorKbar\!^{*0} \to)K\pi$ branching
ratios cannot give an effective constraint due to their larger experimental errors, such that
the uncertainties of the Gegenbauer moments increase dramatically in Scenario II.
For a similar reason, the obtained Gegenbauer moments differ from
those of the $K^*$ meson DA in QCD sum rules~\cite{ball98}, and from
$a_{1K^*}^{0}=0.2$ and $a_{2K^*}^{0}=0.5$ chosen
in the PQCD study on the $B_{(s)}\to\psi K\pi$ decays~\cite{Li:2019pzx}.
We state that the fits based on the currently available data cannot
discriminate the two scenarios effectively. As experimental precision increases, we will be able to impose
more stringent constraints on those two-meson DAs.

%-----------------------%

%-----------------------%
\begin{table}[thp]
\caption{$CP$ averaged branching ratios ${\cal B}$ and direct $CP$ asymmetries ${\cal A}_{CP}$
of the $B_{(s)}\to K (\rho \to) \pi \pi$ decays in the PQCD approach. The experimental data for
comparison are quoted from Ref.~\cite{pdg2020}. Those data marked by $\dagger$ are included in the fit.
The theoretical errors are attributed to the variations of the shape parameter $\omega_{B_{(s)}}$ in the
$B_{(s)}$ meson DA and the decay constant $f_{B_{(s)}}$, of the Gegenbauer moments in the two-pion DAs, and
of the hard scale $t$ and the QCD scale $\Lambda_{\rm QCD}$. }
\label{krho}
\begin{center}
\begin{threeparttable}
\setlength{\tabcolsep}{5mm}{
\begin{tabular}{c c c c} %\toprule
 \hline
{Modes}          &\qquad   & Results              &  Data  \\
\hline
  $B^+ \to K^+(\rho^0\to)\pi \pi$    &~~~${\cal B} (10^{-6})$~~~   &$2.91^{+0.68+0.77+1.43}_{-0.60-0.68-0.82}$     &$3.7\pm{0.5}$~\tnote{$\dagger$}~~\\
                                         &${\cal A}_{CP} (\%)$        &$53.5^{+0.4+4.5+11.9}_{-1.4-4.3-15.0}$         &$37\pm{10}$~\tnote{$\dagger$}~~\\

  $B^0 \to K^+(\rho^-\to)\pi \pi$    &${\cal B} (10^{-6})$        &$8.48^{+2.20+1.63+3.87}_{-1.95-1.48-2.51}$     &$7.0\pm{0.9}$~\tnote{$\dagger$}~~\\
                                         &${\cal A}_{CP} (\%)$        &$33.0^{+1.1+5.2+8.9}_{-1.5-4.9-12.1}$          &$20\pm{11}$\\

  $B_s^0 \to K^-(\rho^+\to)\pi \pi$  &${\cal B} (10^{-6})$    &$16.41^{+7.59+0.16+1.10}_{-5.30-0.15-1.31}$    &$-$\\
                                         &${\cal A}_{CP} (\%)$           &$19.4^{+3.6+3.3+3.1}_{-3.2-3.3-2.9}$     &$-$\\

  $B^+ \to K^0(\rho^+\to)\pi \pi$    &${\cal B} (10^{-6})$         &$7.86^{+2.07+1.51+3.68}_{-1.82-1.50-2.31}$     &$7.3^{+1.0}_{-1.2}$~\tnote{$\dagger$}~~\\
                                         &${\cal A}_{CP} (\%)$          &$13.1^{+1.2+1.8+1.5}_{-0.5-2.5-3.6}$     &$-3\pm{15}$\\

  $B^0 \to K^0(\rho^0\to)\pi \pi$    &${\cal B} (10^{-6})$       &$3.76^{+0.95+0.57+0.92}_{-0.81-0.52-0.81}$     &$3.4\pm{1.1}$~\tnote{$\dagger$}~~\\
                                         &${\cal A}_{CP} (\%)$           &$1.4^{+0.6+0.5+2.1}_{-0.5-0.6-3.1}$      &$-4\pm20$\\

  $B_s^0 \to \bar K^0(\rho^0\to)\pi \pi$       &${\cal B} (10^{-6})$       &$0.17^{+0.04+0.02+0.01}_{-0.04-0.02-0.02}$       &$-$\\
                                                   &${\cal A}_{CP} (\%)$          &$-51.0^{+1.1+11.7+26.6}_{-0.6-10.6-13.4}$       &$-$\\
 \hline
%\bottomrule
\end{tabular}}
%\begin{tablenotes}
%\item $\tnote{$\dagger$}$ Data included in the fit.
%\end{tablenotes}
\end{threeparttable}
\end{center}
\end{table}

\begin{table}[]
\caption{Same as Table~\ref{krho} but for the $B_{(s)}\to \pi (\rho \to) \pi \pi$ decays.}
 \label{pirho}
\begin{center}
\begin{threeparttable}
\setlength{\tabcolsep}{5mm}{
\begin{tabular}{c c c c}
 \hline
{Modes}          &\qquad  & Results              &  Data   \\
\hline
  $B^+ \to \pi^+(\rho^0\to)\pi \pi$       &~~~${\cal B} (10^{-6})$~~~      &$5.98^{+1.56+1.46+0.45}_{-1.37-1.31-0.37}$       &$8.3\pm{1.2}$~\tnote{$\dagger$}~~\\
                                                 &${\cal A}_{CP} (\%)$          &$-34.9^{+2.0+5.3+7.3}_{-0.7-4.4-9.6}$      &$0.9\pm{1.9}$\\

  $B^0 \to \pi^+(\rho^-\to)\pi \pi$           &${\cal B} (10^{-6})$        &$5.28^{+2.08+1.56+0.42}_{-1.58-1.44-0.52}$      &$23.0\pm{2.3}$~\tnote{1}~\tnote{$\dagger$}~~\\   %%%%%%%XXXX
                                                    &${\cal A}_{CP} (\%)$         &$-30.6^{+3.4+4.1+4.5}_{-3.5-4.1-5.4}$     &$-8\pm{8}$\\

  $B^0 \to \pi^-(\rho^+\to)\pi \pi$           &${\cal B} (10^{-6})$        &$20.20^{+8.90+0.48+1.30}_{-6.62-0.54-1.04}$ &$~23.0\pm{2.3}$~\tnote{1}~\tnote{$\dagger$}~~\\
                                                  &${\cal A}_{CP} (\%)$           &$9.3^{+1.9+1.7+1.9}_{-1.6-1.7-1.9}$  &$13\pm6$\\

  $B_s^0 \to \pi^+(\rho^-\to)\pi \pi$         &${\cal B} (10^{-6})$         &$0.23^{+0.04+0.03+0.03}_{-0.04-0.05-0.04}$ &$-$\\
                                                  &${\cal A}_{CP} (\%)$           &$-24.3^{+2.0+4.5+8.8}_{-3.8-14.3-6.1}$               &$-$\\

  $B_s^0 \to \pi^-(\rho^+ \to)\pi \pi$         &${\cal B} (10^{-6})$         &$0.12^{+0.01+0.01+0.00}_{-0.05-0.06-0.06}$  &$-$\\
                                                   &${\cal A}_{CP} (\%)$        &$-71.7^{+2.1+12.0+4.8}_{-1.8-5.6-0.7}$&$-$\\

  $B^+ \to \pi^0(\rho^+\to)\pi \pi$           &${\cal B} (10^{-6})$         &$8.50^{+4.25+1.05+0.24}_{-3.04-0.98-0.55}$  &$10.9\pm{1.4}$~\tnote{$\dagger$}~~\\
                                                  &${\cal A}_{CP} (\%)$           &$20.4^{+5.0+4.6+4.7}_{-4.1-4.4-6.4}$               &$2\pm{11}$\\

  $B^0 \to \pi^0(\rho^0\to)\pi \pi$     &${\cal B} (10^{-6})$         &$0.08^{+0.01+0.02+0.05}_{-0.02-0.03-0.05}$  &$2.0\pm{0.5}$ \\
                                             &${\cal A}_{CP} (\%)$           &$20.8^{+6.0+17.0+11.7}_{-4.4-16.5-40.1}$ &$27\pm24$\\

  $B_s^0 \to \pi^0(\rho^0 \to)\pi \pi$         &${\cal B} (10^{-6})$         &$0.14^{+0.03+0.04+0.04}_{-0.03-0.01-0.04}$ &$-$\\
                                                   &${\cal A}_{CP}  (\%)$         &$-47.9^{+5.5+4.8+4.5}_{-3.0-6.6-7.5}$&$-$\\
 \hline
\end{tabular}}
\begin{tablenotes}
\item $\tnote{1}$ Sum of two branching ratios, ${\cal B}(B\to f) + {\cal B}(B\to \bar{f})$.
%\item $\tnote{$\dagger$}$ Data included in the fit.
\end{tablenotes}
\end{threeparttable}
\end{center}
\end{table}

With the fitted Gegenbauer moments in Table~\ref{tab:gen}, we calculate the $CP$ averaged branching
rations ${\cal B}$ and the direct $CP$ asymmetries ${\cal A}_{CP}$ in the LO PQCD formalism, and present the
results in the central columns of  Tables~\ref{krho}-\ref{pikstar}.
The first theoretical uncertainty originates from
the shape parameter $\omega_B=0.40$~GeV or $\omega_{B_s}=0.48$~GeV with 10\% variation,
and the decay constant $f_{B_{(s)}}$. The second one is from the Gegenbauer moments
in the  two-meson DAs. The last one is caused by the variations of the hard scale $t$ from
$0.75t$ to $1.25t$, which characterizes the effect of next-to-leading-order QCD corrections,
and of the QCD scale $\Lambda_{\rm QCD}=0.25\pm0.05$~GeV.
The errors attributed to the CKM matrix elements are tiny and can be ignored safely.
%\Blue{Among the three different kinds of theoretical errors considered in our work,
%one can see that for most decay channels, the uncertainties for the branching ratios caused by
%the non-perturbative input parameters are significant, especially those from the shape parameters $\omega_{B_{(s)}}$ in the $B_{(s)}$ meson DA,
%which amount to $30\%\sim 50\%$ in magnitude.}
%such as the shape parameters $\omega_{B_{(s)}}$ in the $B_{(s)]$ meson DA as well as
%the Gegenbauer moments in the two-meson DAs, which amount to $30\%\sim 50\%$ in magnitude.}
Note that the data for the $B^0 \to \pi^+(\rho^-\to)\pi \pi$ and $B^0 \to \pi^-(\rho^+\to)\pi \pi$
branching ratios in Table~\ref{pirho} represent the sum over these two modes.
It is also the case for the measured $B_s^0 \to K^+(K^{*-}\to)K\pi$ and $B^+ \to \pi^0(K^{*+}\to)K \pi$
branching ratios, and for the measured $B^0 \to \pi^0(K^{*0}\to)K \pi$ and $B_s^0 \to \bar{K}^0(K^{*0}\to)K \pi$
branching ratios in Table~\ref{kkstar}.

One can also assess the uncertainties from the Gegenbauer moments
$a^{\pi}_{2,4},a^{\pi}_{2P(T)}$ and $a^{K}_{(1,2,4)}$ in the pion and kaon DAs.
Taking the $B^+\to \pi^+ (\rho^0\to)\pi\pi$ and $B^0_s\to K^- (K^{*+}\to)K\pi$ branching rations in Scenario II
as examples, we obtain, given the errors in Eq.~(\ref{eq:genpik}),
\begin{eqnarray}
{\cal B}(B^+\to \pi^+ (\rho^0\to)\pi\pi)&=&(5.98^{+0.02}_{-0.02}(a^{\pi}_{2})^{+0.24}_{-0.26}(a^{\pi}_{4})^{+0.07}_{-0.06}(a^{\pi}_{2P})^{+0.01}_{-0.01}(a^{\pi}_{2T}))\times 10^{-6},\nonumber\\
{\cal B}(B^0_s\to K^- (K^{*+}\to)K\pi)&=&(7.72^{+0.16}_{-0.14}(a^{K}_{1})^{+0.34}_{-0.35}(a^{K}_{2})^{+0.13}_{-0.14}(a^{K}_{4}))\times 10^{-6}.
\label{pkerro}
\end{eqnarray}
It is seen that the former (latter) is more sensitive to the variation of the moment $a^{\pi}_{4}$ ($a^{K}_{2}$)
in the twist-2 pion (kaon) DA.
We remark that the total errors, derived by adding the individual ones from the moments
in the pion and kaon DAs in quadrature and associated with the labels $a^{\pi}$ and $a^{K}$ below,
respectively, are minor (less than $5\%$) compared with other uncertainties listed in Tables~\ref{pirho} and~\ref{kkstar}:
\begin{eqnarray}
{\cal B}(B^+\to \pi^+ (\rho^0\to)\pi\pi)&=&(5.98^{+1.56}_{-1.37}(\omega_B,f_B)^{+0.25}_{-0.26}(a^{\pi})^{+1.46}_{-1.31}(a_{\rho})^{+0.45}_{-0.37}(t,\Lambda_{QCD}))\times 10^{-6},\nonumber\\
{\cal B}(B^0_s\to K^- (K^{*+}\to)K\pi)&=&(7.72^{+1.89}_{-1.59}(\omega_B,f_B)^{+0.40}_{-0.40}(a^{K})^{+1.82}_{-1.49}(a_{K^*})^{+3.24}_{-2.69}(t,\Lambda_{QCD}))\times 10^{-6}.\label{pkerrot}
\end{eqnarray}
Therefore, the variation of the Gegenbauer moments
in the pion and kaon DAs has little impact on the determination of the two-meson DAs.

It is found that most of the considered data in Tables~\ref{krho} and \ref{pirho}
are well reproduced, in particular those with higher precision.
Larger deviation from the data is observed in the $B^+ \to \pi^+(\rho^0\to)\pi \pi$ and
$B^+ \to \pi^0(\rho^+\to)\pi \pi$ branching ratios. It is ascribed to the
involved color-suppressed tree contributions, which receive sizable
next-to-leading-order corrections. The observables removed from the fit
are also predicted in the LO PQCD formalism, and compared with the data in Tables~\ref{krho} and \ref{pirho}.
Our prediction for the $B^0 \to \pi^0(\rho^0\to)\pi \pi$ branching ratio, which suffers significant
subleading corrections as stated before, is still below the data, similar to that derived in the framework for
two-body decays. Most of the ${\cal A}_{CP}$ data for the $B_{(s)} \to P(\rho\to)\pi \pi$ decays with $P=\pi,K$
are not yet precise enough. We mention that ${\cal A}_{CP}$ in the $B^+ \to \pi^+\rho^0$ mode has been
predicted to be large and negative in most QCD approaches~\cite{Cheng:2020hyj,prd95056008}, including the
present analysis on three-body decays as shown in Table~\ref{pirho}. However, its data are as small as
$0.009\pm0.019$~\cite{pdg2020}. Both the theoretical and experimental errors need to be reduced
greatly in order to tell whether the discrepancy really stands as a puzzle.

\begin{table}[thb]
\caption{Same as Table~\ref{krho} but for the $B_{(s)}\to P (\phi \to) KK$ decays with $P=\pi, K$.}
 \label{kk}
\begin{center}
\begin{threeparttable}
\setlength{\tabcolsep}{5mm}{
\begin{tabular}{l c c c}
 \hline
{Modes}          &\qquad  & Results              &  Data   \\
\hline

  $B^+ \to K^+(\phi \to)KK$       &~~~${\cal B} (10^{-6})$~~~     &$8.46^{+3.57+0.41+2.65}_{-2.70-0.45-1.95}$    &$8.8^{+0.7}_{-0.6}$~\tnote{$\dagger$}~~\\
                                 &${\cal A}_{CP} (\%)$             &$1.4^{+0.8+0.1+0.0}_{-0.3-1.7-0.8}$                   &$2.4\pm2.8$\\
  $B^0 \to K^0(\phi \to)K K$      &~~~${\cal B} (10^{-6})$~~~     &$7.82^{+3.18+0.40+2.40}_{-2.50-0.19-1.71}$    &$7.3\pm0.7$~\tnote{$\dagger$}~~\\
                                 &${\cal A}_{CP} (\%)$             &$0$               &$1\pm14$\\

  $B_s^0 \to \bar K^0(\phi \to)K K$       &~~~${\cal B} (10^{-8})$~~~     &$3.52^{+1.30+1.50+2.30}_{-0.64-0.02-1.27}$    &$-$\\
                                 &${\cal A}_{CP} (\%)$                     &$0$                 &$-$\\

 $B^+ \to \pi^+(\phi \to)K K$     &~~~${\cal B} (10^{-8})$~~~     &$1.15^{+0.46+0.02+0.34}_{-0.33-0.20-0.28}$    &$3.2\pm1.5$\\
                                    &${\cal A}_{CP} (\%)$           &$0$                &$10\pm50$\\

  $B^0 \to \pi^0(\phi \to)K K$         &~~~${\cal B} (10^{-9})$~~~     &$5.32^{+2.21+0.14+1.61}_{-1.53-0.91-1.27}$    &$~~<15~~$\\
                                 &${\cal A}_{CP} (\%)$                    &$0$               &$-$\\

  $B_s^0 \to \pi^0(\phi \to)K K$        &~~~${\cal B} (10^{-7})$~~~     &$1.06^{+0.41+0.15+0.07}_{-0.34-0.20-0.14}$   &$-$\\
                                 &${\cal A}_{CP} (\%)$                    &$27.3^{+1.1+3.2+3.5}_{-1.0-1.4-5.8}$                  &$-$\\
 \hline
\end{tabular}}
%\begin{tablenotes}
%\item $\tnote{$\dagger$}$ Data used in the fit.
%\end{tablenotes}
\end{threeparttable}
\end{center}
\end{table}

Both the  $B\to K(\phi \to)KK$ data considered in the fit are well reproduced with a single
Gegenbauer moment $a^0_{2\phi}$ as indicated in Table~\ref{kk}. Our predictions for the branching
ratios and direct $CP$ asymmetries excluded in the fit, mainly associated with $B_s$ meson decays,
can be confronted by more precise data in the future. All the available ${\cal A}_{CP}$ data for the
$B \to P(\phi\to) KK$ decays with $P=\pi,K$ have large errors. The central
values of the prediction and the data for the $B^+ \to \pi^+(\phi \to)K K$ branching ratio are
different, but still agree with each other within uncertainties.

\begin{table}[]
\caption{Same as Table~\ref{krho} but for the $B_{(s)}\to K (K^* \to) K \pi$ decays.}
\label{kkstar}
\begin{center}
\begin{threeparttable}
\setlength{\tabcolsep}{5mm}{
\begin{tabular}{l c c c c}
 \hline
% \multicolumn{1}{c}{}&\multicolumn{1}{c}{}&\multicolumn{2}{c}{ Results } &\multicolumn{1}{c}{} \\
{Modes}          &\qquad  & Results (Scenario I)  & Results (Scenario II)            &  Data   \\
\hline
 $B^+ \to K^+(\bar{K}^{*0} \to)K\pi$         &~~~${\cal B} (10^{-6})$~~~     &$0.55^{+0.14+0.04+0.20}_{-0.13-0.06-0.14}$   &$0.56^{+0.17+0.10+0.15}_{-0.13-0.06-0.13}$          &$0.59\pm0.08$~\tnote{$\dagger$}~~\\
                                             &${\cal A}_{CP} (\%)$             &$46.3^{+1.0+10.9+2.8}_{-0.3-4.7-3.9}$   &$63.8^{+1.1+2.0+3.4}_{-3.2-8.5-23.5}$         &$12\pm10$\\

  $B^0 \to K^+(K^{*-} \to)K\pi$              &~~~${\cal B} (10^{-6})$~~~     &$0.27^{+0.05+0.05+0.04}_{-0.05-0.06-0.03}$   &$0.25^{+0.01+0.09+0.01}_{-0.01-0.03-0.01}$          &$<0.4$~\tnote{1}\\
                                             &${\cal A}_{CP} (\%)$            &$19.8^{+0.5+2.1+13.4}_{-3.6-2.1-7.5}$   &$20.2^{+7.1+10.6+16.9}_{-0.0-1.6-0.0}$                 &$-$\\

  $B^0 \to K^-(K^{*+} \to)K\pi$              &~~~${\cal B} (10^{-6})$~~~     &$0.09^{+0.01+0.01+0.04}_{-0.02-0.01-0.03}$   &$0.11^{+0.02+0.01+0.03}_{-0.06-0.02-0.02}$        &$<0.4$~\tnote{1}\\
                                             &${\cal A}_{CP} (\%)$          &$-5.2^{+12.3+9.4+30.4}_{-15.5-11.4-0.0}$   &$33.8^{+13.4+16.4+9.4}_{-0.0-14.4-0.0}$              &$-$\\

  $B_s^0 \to K^+(K^{*-}\to)K\pi$             &~~~${\cal B} (10^{-6})$~~~     &$15.15^{+2.78+1.90+7.29}_{-2.53-1.72-4.61}$    &$9.89^{+1.92+2.93+5.64}_{-1.66-1.90-4.16}$           &$(19\pm5)$~\tnote{1}~\tnote{$\dagger$}~~\\
                                             &${\cal A}_{CP} (\%)$             &$42.1^{+4.5+2.4+5.5}_{-5.3-3.6-6.9}$     &$6.1^{+0.4+8.8+7.0}_{-1.3-11.0-10.4}$             &$-$\\

  $B_s^0 \to K^-(K^{*+}\to)K \pi$            &~~~${\cal B} (10^{-6})$~~~     &$10.22^{+1.97+1.27+4.51}_{-1.73-1.24-2.72}$      &$7.72^{+1.88+1.82+3.24}_{-1.59-1.49-2.69}$            &$(19\pm5)$~\tnote{1}~\tnote{$\dagger$}~~\\
                                             &${\cal A}_{CP} (\%)$            &$-34.8^{+3.0+1.5+7.5}_{-2.3-0.6-6.6}$    &$-24.0^{+1.5+6.1+11.4}_{-0.3-4.1-6.1}$               &$-$\\

  $B^+ \to \bar{K}^0(K^{*+}\to)K \pi$        &~~~${\cal B} (10^{-6})$~~~     &$0.31^{+0.06+0.07+0.16}_{-0.05-0.04-0.09}$    &$0.19^{+0.06+0.06+0.11}_{-0.05-0.07-0.05}$               &$-$\\
                                             &${\cal A}_{CP} (\%)$            &$-13.6^{+2.5+2.0+5.7}_{-1.0-3.5-7.9}$    &$-22.7^{+13.3+20.7+7.5}_{-0.0-18.4-7.3}$             &$-$\\

  $B^0 \to K^0(\bar{K}^{*0}\to)K \pi$        &~~~${\cal B} (10^{-6})$~~~     &$0.44^{+0.14+0.04+0.15}_{-0.11+0.03+0.11}$    &$0.38^{+0.13+0.05+0.11}_{-0.11-0.04-0.11}$          &$<0.96$~\tnote{1}\\
                                             &${\cal A}_{CP} (\%)$           &$0$       &$0$                     &$-$\\

  $B^0 \to \bar{K}^0(K^{*0}\to)K \pi$         &~~~${\cal B} (10^{-6})$~~~     &$0.44^{+0.08+0.06+0.22}_{-0.08+0.07+0.15}$   &$0.30^{+0.07+0.08+0.16}_{-0.05-0.02-0.12}$           &$<0.96$~\tnote{1}\\
                                              &${\cal A}_{CP} (\%)$             &$0$       &$0$                    &$-$\\

  $B_s^0 \to K^0(\bar{K}^{*0}\to)K \pi$         &~~~${\cal B} (10^{-6})$~~~     &$14.06^{+2.54+1.89+6.88}_{-2.30-1.70-4.48}$   &$8.84^{+1.66+2.77+5.31}_{-1.46-1.98-3.54}$          &$(20\pm6)$~\tnote{1}~\tnote{$\dagger$}~~\\
                                                &${\cal A}_{CP} (\%)$             &$0$     &$0$                          &$-$\\

  $B_s^0 \to \bar{K}^0(K^{*0}\to)K \pi$         &~~~${\cal B} (10^{-6})$~~~     &$10.39^{+2.01+1.18+5.58}_{-1.78-1.17-2.86}$     &$7.92^{+1.95+1.63+3.46}_{-1.64-1.36-2.85}$        &$(20\pm6)$~\tnote{1}~\tnote{$\dagger$}~~\\
                                                &${\cal A}_{CP} (\%)$                 &$0$      &$0$                       &$-$\\

 \hline
\end{tabular}}
%\begin{tablenotes}
%\item $\tnote{1}$ Sum of two branching ratios, ${\cal B}(B_{(s)}\to f) + {\cal B}(B_{(s)}\to \bar{f})$.
%\item $\tnote{$\dagger$}$ Data used in the fit.
%\end{tablenotes}
\end{threeparttable}
\end{center}
\end{table}

\begin{table}[thb]
\caption{Same as Table~\ref{krho} but for the $B_{(s)}\to \pi (K^* \to) K \pi$ decays.}
 \label{pikstar}
\begin{center}
\begin{threeparttable}
\setlength{\tabcolsep}{5mm}{
\begin{tabular}{ l c c c c}
 \hline
% \multicolumn{1}{c}{}&\multicolumn{1}{c}{}&\multicolumn{2}{c}{ Results } &\multicolumn{1}{c}{} \\
{Modes}          &\qquad  & Results (Scenario I)  & Results (Scenario II)                &  Data   \\
\hline
  $B^+ \to \pi^+(K^{*0}\to)K \pi$     &~~~${\cal B} (10^{-6})$~~~     &$7.17^{+1.56+0.64+3.46}_{-1.37-0.62-2.23}$  &$8.19^{+2.14+0.94+2.74}_{-1.77-0.66-1.93}$  &$~~10.1\pm{0.8}~~$\\
                                    &${\cal A}_{CP} (\%)$            &$-5.4^{+0.5+0.8+2.1}_{-0.2-0.3-0.8}$          &$-4.5^{+0.5+1.1+2.7}_{-0.6-1.4-1.2}$             &$-4\pm9$\\

  $B^0 \to \pi^-(K^{*+}\to)K \pi$         &~~~${\cal B} (10^{-6})$~~~     &$7.47^{+1.60+0.72+3.29}_{-1.41-0.71-2.06}$  &$7.61^{+1.83+0.92+2.40}_{-1.61-0.65-1.78}$   &$7.5\pm{0.4}$~\tnote{$\dagger$}~~\\
                                                   &${\cal A}_{CP} (\%)$    &$-52.9^{+3.1+0.7+9.3}_{-1.8-1.0-7.0}$    &$-32.3^{+0.7+10.3+7.9}_{-0.2-8.4-6.3}$           &$-27\pm{4}$\\

  $B_s^0 \to \pi^+(K^{*-}\to)K\pi$       &~~~${\cal B} (10^{-6})$~~~     &$12.13^{+4.66+1.36+0.92}_{-3.55-1.29-0.75}$ &$5.52^{+2.22+2.09+0.41}_{-1.66-1.85-0.41}$   &$~~2.9\pm{1.1}~~$\\
                                 &${\cal A}_{CP} (\%)$                     &$-32.8^{+4.1+2.7+4.2}_{-4.8-3.2-5.5}$  &$-30.6^{+4.1+6.5+8.2}_{-4.5-6.9-8.9}$                 &$-$\\

  $B^+ \to \pi^0(K^{*+}\to)K \pi$      &~~~${\cal B} (10^{-6})$~~~     &$4.71^{+1.18+0.39+1.92}_{-0.98-0.38-1.30}$ &$5.62^{+1.54+0.62+1.55}_{-1.28-0.49-1.11}$   &$6.8\pm{0.9}$~\tnote{$\dagger$}~~\\
                                 &${\cal A}_{CP} (\%)$                  &$-36.2^{+1.6+0.1+7.4}_{-1.0-0.4-8.2}$  &$-19.1^{+2.4+6.6+4.6}_{-1.9-5.4-6.0}$             &$-39\pm{21}$\\

  $B^0 \to \pi^0(K^{*0}\to)K \pi$       &~~~${\cal B} (10^{-6})$~~~     &$2.99^{+0.59+0.33+1.49}_{-0.55-0.33-0.89}$ &$2.55^{+0.57+0.36+1.06}_{-0.49-0.19-0.74}$   &$3.3\pm{0.6}$~\tnote{$\dagger$}~~\\
                                 &${\cal A}_{CP} (\%)$                 &$-11.6^{+1.0+0.49+5.0}_{-1.2-0.2-1.0}$ &$-11.8^{+1.2+4.3+4.3}_{-1.2-1.7-0.2}$                &$-15\pm{13}$\\

  $B_s^0 \to \pi^0(\bar{K}^{*0}\to)K \pi$        &~~~${\cal B} (10^{-6})$~~~     &$0.20^{+0.03+0.01+0.06}_{-0.04-0.02-0.05}$ &$0.12^{+0.02+0.02+0.03}_{-0.04-0.03-0.03}$  &$-$\\
                                 &${\cal A}_{CP} (\%)$                           &$-70.6^{+6.7+13.2+23.5}_{-6.7-2.8-15.1}$  &$-50.4^{+3.1+22.7+15.1}_{-2.6-12.4-14.1}$                &$-$\\
 \hline
\end{tabular}}
%\begin{tablenotes}
%\item $\tnote{$\dagger$}$ Data used in the fit.
%\end{tablenotes}
\end{threeparttable}
\end{center}
\end{table}

Overall speaking, Scenario II reproduces the considered $B_{(s)} \to P(K^{*}\to)K\pi$
data with $P=\pi,K$ slightly better than Scenario I does as seen
in Tables~\ref{kkstar} and \ref{pikstar}. The $B_s \to P(K^{*}\to)K\pi$
branching ratios differ between the two scenarios more than the $B \to P(K^{*} \to)K\pi$ branching
ratios do. This feature is understandable, because the former involve the $B_s \to (K^{*}\to) K\pi$
transition form factors, which are more sensitive to the variation of the Gegenbauer moments in the
$K\pi$ DA. Hence, more precise $B_s \to P(K^{*}\to)K\pi$ data are crucial for fixing the $K\pi$ DAs.
The direct $CP$ asymmetries ${\cal A}_{CP}$ in some $B_{(s)} \to P(K^{*}\to)K\pi$ modes depend on
the chosen scenarios strongly, implying that more accurate $K\pi$ DAs are necessary for
predicting these observables unambiguously. The central value of the
predicted $B_s^0 \to \pi^+(K^{*-}\to)K\pi$ branching ratio in Scenario II, which is already much
lower than in Scenario I, remains above the data. It deserves more thorough theoretical and
experimental investigations. Similarly, most of the ${\cal A}_{CP}$ data for the $B_{(s)} \to P(K^{*}\to)K\pi$
decays have substantial uncertainties so far, so it is not yet possible to make a meaningful
comparison with our results.

A remark is in order. The twist-2 DAs $\phi_K^{A}$ and $\phi^0_{K\pi}$ in Eqs.~(\ref{eq:pkda}) and (\ref{eq:pikldt}),
respectively, are expanded up to the fourth-order Gegenbauer polynomial without the third-order term,
which exists in general. We find that the $SU(3)$ breaking effects in the considered decays
can be well accounted for by the first-order term alone. That is, even the third-order term is
included into the fit, its value turns out to be small, and  does not modify the fit outcomes
much. Taking the eight $B_{(s)} \to P(K^*\to)K\pi$ ($P=\pi,K$) decays as examples,
we perform the global fit with $a_{3K^*}^0$ being included, and obtain
the Gegenbauer moments of the $K\pi$ twist-2 DA
\begin{eqnarray}
a_{1K^*}^{0}=0.37\pm0.60,\;\;\;\;  a_{2K^*}^{0}=1.19\pm0.10,\;\;\;\;  a_{3K^*}^{0}=-0.04\pm0.36,
\label{tab:kpi123}
\end{eqnarray}
%in Table~\ref{tab:kpi123}
and the branching ratios in Table~\ref{tab:resultkpi123}. It is seen that the central value of
$a_{1K^*}^0$ increases only a bit with an enlarged uncertainty
and $a_{2K^*}^0$  stays the same, compared with those from Scenario I in Table~\ref{tab:gen}, and
the central value of $a_{3K^*}^0$ is tiny. The corresponding branching ratios  in Table~\ref{tab:resultkpi123} also
change very little, compared with those in Tables~\ref{kkstar} and \ref{pikstar}. The above observations
support that the $SU(3)$ breaking effects in the considered modes can be explained  by the
$a_{1K^*}^0$ term alone under the current data precision. Hence, the neglect of the
$a_{3K^*}^0$ term in this work is justified. Besides, it is not practical to include many
parameters in the fit because of the limited amount of experimental data at present.
For a similar reason, the asymptotic forms of the $K\pi$ twist-3 DAs $\phi_{K\pi}^{s,t} $ are adopted
in our analysis. The same argument applies to the expansion of the kaon DAs
in Eq.~(\ref{eq:pkda}), where the higher moments responsible for $SU(3)$ symmetry breaking effects are also absent.
We will explore the impact of these neglected Gegenbauer polynomials
systematically in the future, when more experimental data with improved precision are available.

%\begin{table}[!htbh]
%\caption{Fitted Gegenbauer moments $a^0_{1K^*}$, $a^0_{2K^*}$, $a^0_{3K^*}$ for twist-2 $K\pi$ DAs.}
%\label{tab:kpi123}
%\begin{center}
%\setlength{\tabcolsep}{3mm}{
%\begin{tabular}{lcccccccccccccc}
%		\hline
%&$a_{1K^*}^{0}$    &$a_{2K^*}^{0}$  &$a_{3K^*}^{0}$            \\ \hline
%\ fit \     &$0.37\pm0.60$ & $1.19\pm0.10$ &$-0.04\pm0.36$       \\  \hline
%\end{tabular}}
%\end{center}
%\end{table}
%-----------------------%
\begin{table}[htbp!]
\caption{$CP$ averaged branching ratios ${\cal B} (10^{-6})$ corresponding to the fitted Gegenbauer
moments in Eq.~(\ref{tab:kpi123}), compared with the data~\cite{pdg2020}.
For simplicity, only the theoretical errors from the Gegenbauer moments are
presented. }
\label{tab:resultkpi123}
\begin{center}
\begin{threeparttable}
\setlength{\tabcolsep}{8mm}{
\begin{tabular}{lcc}
		\hline
		{Modes}                                   &Results &Data\\
	\hline
$B^+\to K^+(\bar K^{*0}\to)K\pi$        &$0.59\pm 0.08$  &$3.7\pm0.5$~\tnote{$\dagger$}~~     \\
$B^0\to \pi^-(K^{*+}\to)K\pi$        &$7.51\pm 0.34$   &$7.5\pm0.4$~\tnote{$\dagger$}~~     \\
$B^+\to \pi^0(K^{*+}\to)K\pi$       &$4.75\pm 0.37$   &$6.8\pm0.9$~\tnote{$\dagger$}~~   \\
$B^0\to \pi^0(K^{*0}\to)K\pi$       &$2.91\pm 0.40$ &$3.3\pm0.6$~\tnote{$\dagger$}~~  \\
$B_s^0\to K^+(K^{*-}\to)K\pi+c.c$        &$25.40\pm 1.60$   &$19\pm5$~\tnote{$\dagger$}~~  \\
$B_s^0\to K^0(\bar K^{*0}\to)K\pi+c.c$      &$24.60\pm 1.50$ &$20\pm6$~\tnote{$\dagger$}~~  \\
\hline
\end{tabular}}
%\begin{tablenotes}
%\item $\tnote{1}$ Sum of two branching ratios, ${\cal B}(B_{(s)}\to f) + {\cal B}(B_{(s)}\to \bar{f})$.
%\item $\tnote{$\dagger$}$ Data used in the fit.
%\end{tablenotes}
\end{threeparttable}
\end{center}
\end{table}
%-----------------------%

It is noticed that the parametrization of the parton momenta in Eqs.~(\ref{mom-B-k}) and (\ref{fg})
introduces the dependence on the light meson mass $m_3$ into the hard kernels and the
Sudakov exponents, as explicitly shown in the Appendix. Since both these factors are perturbative pieces
in a PQCD factorization formula, they should be insensitive to a light scale. Therefore,
we test the sensitivity of our numerical results to this light scale by setting it to zero in the
hard kernels and the Sudakov exponents. The corresponding branching ratios and
direct $CP$ asymmetries for two typical modes, $B^+ \to K^+(\rho^0\to)\pi \pi$ and
$B^0 \to \pi^-(K^{*+}\to)K\pi$ in Scenario II, are presented in Table~\ref{krho1}.
The neglect of the kaon mass for the former mode causes about 10\% variation
in the branching ratio and the direct $CP$ asymmetry. The quantities associated with the
latter mode are relatively stable with respect to the neglect of the
pion mass as expected. The insensitivity to the light scale confirms that our parametrization for
kinematic variables in three-body $B$ meson decays is reasonable.

%-----------------------%
\begin{table}[thp]
\caption{$CP$ averaged branching ratios and direct $CP$ asymmetries
of the $B^+ \to K^+(\rho^0\to)\pi \pi$ decay and the $B^0 \to \pi^-(K^{*+}\to)K\pi$ decay in
Scenario II with and without the light meson masses in the hard kernels and the Sudakov exponents. The experimental data
are quoted from~\cite{pdg2020}. The sources of the theoretical errors are the same as in Table~\ref{krho}.}
\label{krho1}
\begin{center}
\setlength{\tabcolsep}{5mm}{
\begin{tabular}{lllll}
 \hline
{Modes}          &\qquad   & Results (with light mass)  & Results (without light mass)            &  Data  \\
\hline
  $B^+ \to K^+(\rho^0\to)\pi \pi$    &${\cal B} (10^{-6})$
  &$2.91^{+0.68+0.77+1.43}_{-0.60-0.68-0.82}$ &$2.51^{+0.56+0.71+1.34}_{-0.52-0.53-0.80}$    &$3.7\pm{0.5}$\\
  &${\cal A}_{CP} (\%)$     &$53.5^{+0.4+4.5+11.9}_{-1.4-4.3-15.0}$
  &$58.5^{+0.0+4.4+11.9}_{-1.9-6.6-17.1}$       &$37\pm{10}$\\

  $B^0 \to \pi^-(K^{*+}\to)K\pi$    &${\cal B} (10^{-6})$  &$7.61^{+1.83+0.92+2.40}_{-1.61-0.65-1.78}$
  &$7.66^{+1.84+0.95+2.43}_{-1.60-0.64-2.06}$   &$7.5\pm{0.4}$\\
  &${\cal A}_{CP} (\%)$     &$-32.3^{+0.7+10.3+7.9}_{-0.2-8.4-6.3}$
  &$-32.7^{+0.6+10.4+7.9}_{-0.1-8.4-6.1}$     &$-27\pm{4}$\\
 \hline
%\bottomrule
\end{tabular}}
\end{center}
\end{table}

\subsection{$\omega^2$-dependent Gegenbauer Moments}

We make a more aggressive attempt in this subsection to determine the
dependence of the Gegenbauer moments in the two-meson DAs on the meson
pair invariant mass. As stated in the Introduction, it is unlikely to extract the exact dependence
from current data, so we simply expand the Gegenbauer moments up to the first power in
$\omega^2$, and examine whether the additional linear terms can be constrained effectively in the
global fit. Consider the parametrizations of the di-pion DAs,
\begin{eqnarray}
\phi_{\pi\pi}^0(x,\omega^2)&=&\frac{3F_{\pi\pi}^{\parallel}(\omega^2)}{\sqrt{2N_c}}x(1-x)\left[1
+a^0_{2\rho}(1+c_\rho^0\omega^2)C_2^{3/2}(2x-1)\right] ,\nonumber\\ %\label{eq:pild0}
\phi_{\pi\pi}^{s}(x,\omega^2)&=&\frac{3F_{\pi\pi}^{\perp}(\omega^2)}{2\sqrt{2N_c}}(1-2x)\left[1
+a^s_{2\rho}(1+c_\rho^s\omega^2)(10x^2-10x+1)\right] ,\nonumber\\%\label{eq:pilds}
\phi_{\pi\pi}^t(x,\omega^2)&=&\frac{3F_{\pi\pi}^{\perp}(\omega^2)}{2\sqrt{2N_c}}(1-2x)^2\left[1
+a^t_{2\rho}(1+c_\rho^t\omega^2)C_2^{3/2}(2x-1)\right] ,
\end{eqnarray}
with the free parameters $a^{0,s,t}_{2\rho}$ and $c_\rho^{0,s,t}$.
The above parametrization follows the power series for the $\omega^2$-dependent
Gegenbauer moments derived in Ref.~\cite{MP}.

%-----------------------%
\begin{table}
\caption{Fitted parameters for the $\omega^2$-dependent Gegenbauer moments in the twist-2 and twist-3 $\pi\pi$ DAs.}
\label{tab:genc0st}
\begin{center}
\setlength{\tabcolsep}{4mm}{
\begin{tabular}{lcccccc}
%\begin{tabular}{lcccccc}
		\hline
		 &$a^0_{2\rho}$             &$a^s_{2\rho}$                &$a^t_{2\rho}$   &$c^0_{\rho}$ (GeV$^{-2}$)
		 &$c^s_{\rho}$ (GeV$^{-2}$) &$c^t_{\rho}$ (GeV$^{-2}$)        \\ \hline
\ fit \  &$-0.45\pm0.29$            &$1.12\pm0.33$                &$-0.43\pm0.11$    &$-0.44\pm0.93$
& $-1.42\pm0.42$ &$-0.03\pm0.32$              \\ \hline

\end{tabular}}
\end{center}
\end{table}
%-----------------------%

The global fit to the same set of $B_{(s)}\to P(\rho\to)\pi\pi$ data with $P=\pi,K$
leads to the outcomes in Table~\ref{tab:genc0st} with a smaller $\chi^2 / d.o.f.=0.51$, which
are not difficult to understand: varying $\omega^2$ around the $\rho$ resonance
in its width window, we find that the values of $a^{0,s,t}_{2\rho}(1+c_\rho^{0,s,t}\omega^2)$
are in fact consistent with the corresponding ones in Table~\ref{tab:gen}. The consistency
is particularly obvious for  $a^{t}_{2\rho}(1+c_\rho^{t}\omega^2)$ with the tiny coefficient $c_\rho^{t}$.
It is observed from Table~\ref{tab:genc0st} that the parameters for the twist-3 DA $\phi_{\pi\pi}^{s}$,
which gives sizable contributions to branching ratios,
can be constrained effectively by the current data. It suggests that the determination of
the $\omega^2$-dependent Gegenbauer moments is promising, when more precise data are available in the
future. Because our purpose is to demonstrate the potential to extract the $\omega^2$ dependence
of the Gegenbauer moments, we will not work on the $K\pi$ and $KK$ DAs.
The effect of including the $\omega^2$ dependence of the Gegenbauer moments is
similar to that of introducing more parameters. That is, the fit quality is improved with a lower
$\chi^2 / d.o.f.$ at the cost of larger uncertainties for fit results as shown in Table~\ref{tab:fitproces}.
For example, the reproduced branching ratios for the $B^+\to K^+(\rho^0\to)\pi\pi$ and
$B^+\to \pi^+(\rho^0\to)\pi\pi$ decays get closer to the data, which have relatively higher
precision. However, the uncertainty caused by the variation of the di-pion DAs is amplified
compared to the second source of errors in Table~\ref{krho}.

%-----------------------%
\begin{table}[htbp!]
\caption{
$CP$ averaged branching ratios and direct $CP$ asymmetries derived
from the fitted Gegenbauer moments in Table~\ref{tab:genc0st}, and compared with data~\cite{pdg2020}.
For simplicity, only the theoretical errors from the Gegenbauer moments are
presented. }
\label{tab:fitproces}
\begin{center}
\begin{threeparttable}
\setlength{\tabcolsep}{8mm}{
\begin{tabular}{lccc}
		\hline
		{Modes}      &\qquad                              &Results &Data\\
	\hline
$B^+\to K^+(\rho^0\to)\pi\pi$    &${\cal B} (10^{-6})$     &$3.12^{+1.81}_{-1.14}$  &$3.7\pm0.5$~\tnote{$\dagger$}~~     \\
&${\cal A}_{CP} (\%)$          &$37.9^{+10.6}_{-11.9}$     &$37\pm 10$~\tnote{$\dagger$}~~\\
$B^+\to K^0(\rho^+\to)\pi\pi$     &${\cal B} (10^{-6})$    &$8.66^{+3.24}_{-1.99}$   &$7.3\pm1.2$~\tnote{$\dagger$}~~  \\
&${\cal A}_{CP} (\%)$          &$17.8^{+2.6}_{-1.1}$ &$-3\pm15$\\
$B^0\to K^+(\rho^-\to)\pi\pi$    &${\cal B} (10^{-6})$     &$8.22^{+3.21}_{-1.93}$ &$7.0\pm0.9$~\tnote{$\dagger$}~~  \\
&${\cal A}_{CP} (\%)$     &$18.9^{+7.8}_{-8.7}$   &$20\pm11$\\
$B^0\to K^0(\rho^0\to)\pi\pi$    &${\cal B} (10^{-6})$     &$2.88^{+1.19}_{-0.72}$   &$3.4\pm1.1$~\tnote{$\dagger$}~~     \\
&${\cal A}_{CP} (\%)$   &$1.9^{+1.8}_{-0.8}$  &$-4\pm20$\\
$B^+\to \pi^+(\rho^0\to)\pi\pi$   &${\cal B} (10^{-6})$    &$7.69^{+2.67}_{-1.65}$   &$8.3\pm1.2$~\tnote{$\dagger$}~~   \\
&${\cal A}_{CP} (\%)$  &$-17.2^{+9.2}_{-4.2}$ &$0.9\pm1.9$\\
$B^+\to \pi^0(\rho^+\to)\pi\pi$   &${\cal B} (10^{-6})$    &$10.14^{+5.18}_{-3.89}$ &$10.9\pm1.4$~\tnote{$\dagger$}~~  \\
&${\cal A}_{CP} (\%)$  &$5.6^{+13.4}_{-14.8}$ &$2\pm11$\\
$B^0\to \pi^-(\rho^+\to)\pi\pi$    &${\cal B} (10^{-6})$   &$24.49^{+2.29}_{-1.72}$~\tnote{1} &$23.0\pm2.3$~\tnote{1}~\tnote{$\dagger$}~~  \\
&${\cal A}_{CP} (\%)$  &$3.8^{+5.1}_{-5.3}$ &$13\pm6$\\
$B^0\to\pi^+(\rho^-\to)\pi\pi$     &${\cal B} (10^{-6})$   &$24.49^{+2.29}_{-1.72}$~\tnote{1} &$23.0\pm2.3$~\tnote{1}~\tnote{$\dagger$}~~   \\
&${\cal A}_{CP} (\%)$  &$-16.4^{+11.8}_{-10.1}$ &$-8\pm8$\\
\hline
\end{tabular}}
%\begin{tablenotes}
%\item $\tnote{1}$ Sum of two branching ratios, ${\cal B}(B_{(s)}\to f) + {\cal B}(B_{(s)}\to \bar{f})$.
%\item $\tnote{$\dagger$}$ Data used in the fit.
%\end{tablenotes}
\end{threeparttable}
\end{center}
\end{table}
%-----------------------%

\section{CONCLUSION}

In this work we have performed a global fit of the Gegenbauer moments in two-meson DAs to measured
branching ratios and direct $CP$ asymmetries in the three-body hadronic $B$ meson decays
$B\to VP_3\to P_1P_2P_3$ with $V=\rho,\phi, K^*$ and $P_3=\pi,K$ in the LO PQCD approach.
Two-meson DAs, collecting both nonresonant and multi-resonance contributions, serve as
crucial nonperturbative ingredients of factorization theorems for the above decays.
The Gegenbauer moments of the pion and kaon DAs determined in the LO global analysis
of two-body hadronic $B$ meson decays have been input
for theoretical consistency. To facilitate the numerical study,
we have constructed a Gegenbauer-moment-independent database, via which a decay amplitude
is decomposed into a linear combination of the relevant Gegenbauer moments in the two-meson
DAs. It was noticed that the fitted Gegenbauer moments differ from those associated with
an intermediate resonance which decays into the meson pair, and from those adopted in previous PQCD
calculations. This observation indicates that
the Gegenbauer moments of a two-meson DA cannot be inferred from sum-rule results
for an intermediate resonance, and their global determination is essential.

We have examined two scenarios for the determination of the $K\pi$ DAs in order to
check the convergence of the Gegenbauer expansion, and the sensitivity of the fitted
observables to our setup. It was found that the Gegenbauer expansion is improved by increasing
the number of Gegenbauer moments at the cost of large uncertainties for fit outcomes,
and that the branching ratios of $B_s$ meson decays and direct $CP$ asymmetries in
some modes are more sensitive to the chosen scenarios. Hence, more accurate $K\pi$ DAs
are necessary for predicting these quantities unambiguously. We state that
our fits have not been able to discriminate the two scenarios effectively.
We have also explored the potential to fix the
dependence of the Gegenbauer moments on the meson-pair invariant mass, and confirmed that
at least the parameter for the twist-3 DA $\phi_{\pi\pi}^{s}$ can be constrained to some extent by the
current data. Therefore, the determination of the dependence on the meson-pair invariant mass
is promising, when data become more precise.

We mention that the three-body charmless hadronic $B$ meson decays included in this work
have been studied in Refs.~\cite{epjc80394,epjc80517,epjc7937,prd95056008} in a scattered manner.
The improvements compared to the earlier studies contain
(1) the partonic kinematic variables have been refined to take into account finite masses of final-state
mesons, such that the $SU(3)$ symmetry breaking effects in the decays can be evaluated more precisely; and
(2) the Gegenbauer moments in the two-meson DAs have been determined in a global analysis for the first time,
which are valuable for future applications of the PQCD framework to multi-body $B$ meson decays; (3) the dependence
of the Gegenbauer moments on the meson-pair invariant mass has been probed for the first time. Because of
(1), the numerous hard kernels involved in the various modes need to be modified, which have been presented,
together with the factorization formulas for the decay amplitudes, in the Appendix. The refined
partonic kinematics is general enough for its extension to multi-body $B$ meson decays into arbitrary massive
final states. For (2), we remind that different Gegenbauer moments for the $K\pi$ DAs were taken
in the previous scattered studies, such as Refs.~\cite{epjc80517} and~\cite{epjc7937}, so our work facilitates a
consistent understanding of multi-body $B$ meson decays. We have shown that the preferred central value of,
for instance, the Gegenbauer moment $a_{1K^*}^{0}$ is 0.31, instead of 0.2 in~\cite{epjc80517}
or 0.05 in~\cite{epjc7937} (but note the large theoretical uncertainties).

It has been demonstrated that most of the data considered in the fit are well reproduced,
namely, the fit quality is satisfactory. It implies that the two-meson DAs presented in
this paper are ready for applications to other multi-body hadronic $B$ meson decays
involving the same meson pairs. With the obtained Gegenbauer moments, we have made
predictions for those observables, whose data were excluded in the fit because of
their substantial experimental errors or significant subleading contributions to the
corresponding factorization formulas. Except the $B_s^0 \to \pi^+(K^{*-}\to)K\pi$ branching
ratio, our predictions agree with the data within uncertainties in the former case.
Since our results were still derived in the LO PQCD approach, the data in the latter
case remain unexplained, and deserve more through investigations. As pointed out
before, the precision of the extracted two-meson DAs can be improved systematically,
when higher-order and/or higher-power corrections to three-body hadronic $B$ meson decays
are taken into account in our formalism. At the same time, more precise measurements are urged,
especially those of $CP$ asymmetries. These efforts will
strengthen the constraint on the Gegenbauer moments and sharpen the confrontation
between theoretical predictions and experimental data.

\begin{acknowledgments}
We thank W.F. Wang for helpful discussions.
This work is supported in part by ``Fundamental Research Funds for Central Universities''
under Grant No.~KJQN202144 and the National Natural Science Foundation of China under
Grant Nos.~12005103, 12075086, 11947013, 11947215 and U2032102, and by MOST of R.O.C. under Grant No. MOST-107-2119-M-001-035-MY3.
YL is also supported by the Natural Science Foundation of Jiangsu Province under
Grant No.~BK20190508 and the Research Start-up Funds of Nanjing Agricultural University.
DCY is supported by the Natural Science Foundation of Jiangsu Province under Grant No.~BK20200980.
ZR is supported in part by the Natural Science Foundation of Hebei Province under Grant Nos.~A2019209449 and A2021209002.
\end{acknowledgments}

\appendix\label{sec:ap}
\section{Decay amplitudes}

In this Appendix we present the PQCD factorization formulas for the amplitudes of
the considered three-body charmless hadronic $B$ meson decays.
%$B_{(s)} \to P (\rho\to)\pi\pi$, $B_{(s)} \to P (K^*\to)K\pi$ and $B_{(s)} \to P (\phi\to)KK$ with $P=\pi, K$.
We first decompose various decay amplitudes $\cal A$ in terms of the factorizable emission
(annihilation) contributions $F_{e(a)V}$ and the nonfactorizable emission (annihilation) contributions
$M_{e(a)V}$ for the intermediate vector mesons $V=\rho,K^*,\phi$ from Fig.~\ref{fig:fig1}, and
the similar ones $F_{e(a)P}$ and $M_{e(a)P}$ for the bachelor mesons $P=\pi,K$ from Fig.~\ref{fig:fig2}.
These contributions are further labelled by the superscripts $LL$, $LR$ and $SP$ corresponding to the
$(V-A)(V-A)$, $(V-A)(V+A)$ and $(S-P)(S+P)$ operators, respectively:
\begin{enumerate}
\item[$\bullet$]  $B_{(s)} \to K (\rho \to) \pi \pi$
\beq
{\cal A}(B^+ \to K^+(\rho^0 \to)\pi\pi) &=& \frac{G_F} {2} \big\{V_{ub}^*V_{us}[(\frac{C_1}{3}+C_2)(F^{LL}_{e\rho}+F^{LL}_{a\rho})+(C_1+\frac{C_2}{3})F^{LL}_{eK}+C_2M^{LL}_{eK}\non
&+&C_1(M^{LL}_{e\rho}+M^{LL}_{a\rho})]-V_{tb}^*V_{ts}[(\frac{C_3}{3}+C_4+\frac{C_9}{3}+C_{10})(F^{LL}_{e\rho}+F^{LL}_{a\rho})\non
&+&(\frac{C_5}{3}+C_6+\frac{C_7}{3}+C_8)(F^{SP}_{e\rho}+F^{SP}_{a\rho})+(C_3+C_9)(M^{LL}_{e\rho}+M^{LL}_{a\rho})\non
&+&(C_5+C_7)(M^{LR}_{e\rho}+M^{LR}_{a\rho})+\frac{3 C_8}{2} M^{SP}_{eK}+\frac{3 C_{10} }{2} M^{LL}_{eK} \non
&+&\frac{3}{2}(C_7+\frac{C_8}{3}+C_9+\frac{C_{10}}{3})F^{LL}_{eK}]\big\} , \label{amp1}
 %------------------------------------------------------------------------------------------------
 \eeq
 \beq
{\cal A}(B^0 \to K^+(\rho^- \to)\pi\pi) &=& \frac{G_F} {\sqrt{2}} \big\{V_{ub}^*V_{us}[(\frac{C_1}{3}+C_2)F^{LL}_{e\rho}+C_1 M^{LL}_{e\rho}]-V_{tb}^*V_{ts}[(C_3\non
&+&C_9)M^{LL}_{e\rho}+(\frac{C_3}{3}+C_4+\frac{C_9}{3}+C_{10})F^{LL}_{e\rho}+(\frac{C_5}{3}+C_6+\frac{C_7}{3}\non
&+&C_8)F^{SP}_{e\rho}+(C_5+C_7)M^{LR}_{e\rho}+(\frac{C_3}{3}+C_4-\frac{1}{2}(\frac{C_9}{3}+C_{10}))F^{LL}_{a\rho}\non &+&(C_3-\frac{C_9}{2})M^{LL}_{a\rho}+(\frac{C_5}{3}+C_6-\frac{1}{2}(\frac{C_7}{3}+C_8))F^{SP}_{a\rho}+(C_5\non
&-&\frac{C_7}{2})M^{LR}_{a\rho}]\big\} ,\label{amp2}
%------------------------------------------------------------------------------------------------
\eeq
\beq
 {\cal A}(B_s^0 \to K^-(\rho^+\to)\pi\pi) &=&  \frac{G_F} {\sqrt{2}} \big\{V_{ub}^*V_{ud}[(\frac{C_1}{3}+C_2)F^{LL}_{eK}+C_1 M^{LL}_{eK}]-V_{tb}^*V_{td}[(C_3+C_9)M^{LL}_{eK}\non
&+&(\frac{C_3}{3}+C_4+\frac{C_9}{3}+C_{10})F^{LL}_{eK}+(C_5+C_7)M^{LR}_{eK}+(\frac{C_3}{3}+C_4\non
&-&\frac{1}{2}(\frac{C_9}{3}+C_{10}))F^{LL}_{aK}+(\frac{C_5}{3}+C_6-\frac{1}{2}(\frac{C_7}{3}+C_8))F^{SP}_{aK}\non
&+&(C_3-\frac{C_9}{2})M^{LL}_{aK}+(C_5-\frac{C_7}{2})M^{LR}_{aK}]\big\} ,\label{amp3}
 %------------------------------------------------------------------------------------------------
 \eeq
 \beq
 {\cal A}(B^+ \to K^0(\rho^+\to)\pi\pi) &=& \frac{G_F} {\sqrt{2}}\big\{V_{ub}^*V_{us}[(\frac{C_1}{3}+C_2)F^{LL}_{a\rho}+C_1 M^{LL}_{a\rho}]-V_{tb}^*V_{ts}[(C_3-\frac{C_9}{2})M^{LL}_{e\rho}\non
&+&(\frac{C_3}{3}+C_4-\frac{1}{2}(\frac{C_9}{3}+C_{10}))F^{LL}_{e\rho}+(\frac{C_5}{3}+C_6-\frac{1}{2}(\frac{C_7}{3}+C_8))F^{SP}_{e\rho}\non
&+&(C_5-\frac{C_7}{2})M^{LR}_{e\rho}+(\frac{C_3}{3}+C_4+\frac{C_9}{3}+C_{10})F^{LL}_{a\rho}+(C_3+C_9)M^{LL}_{a\rho}\non
&+&(\frac{C_5}{3}+C_6+\frac{C_7}{3}+C_8)F^{SP}_{a\rho}+(C_5+C_7)M^{LR}_{a\rho}]\big\} ,\label{amp4}
\eeq
 %------------------------------------------------------------------------------------------------
 \beq
{\cal A}(B^0 \to K^0(\rho^0\to)\pi\pi) &=& \frac{G_F} {2}\big\{V_{ub}^*V_{us}[(C_1+\frac{C_2}{3})F^{LL}_{eK}+C_2 M^{LL}_{eK}]-V_{tb}^*V_{ts}[\frac{3 C_8}{2}M^{SP}_{eK}-(\frac{C_3}{3}\non
&+&C_4-\frac{1}{2}(\frac{C_9}{3}+C_{10}))(F^{LL}_{e\rho}+F^{LL}_{a\rho})-(C_3-\frac{C_9}{2})(M^{LL}_{e\rho}+M^{LL}_{a\rho})\non
&-&(\frac{C_5}{3}+C_6-\frac{1}{2}(\frac{C_7}{3}+C_8))(F^{SP}_{e\rho}+F^{SP}_{a\rho})-(C_5-\frac{C_7}{2})(M^{LR}_{e\rho}\non
&+&M^{LR}_{a\rho})+\frac{3}{2}(C_7+\frac{C_8}{3}+C_9+\frac{C_{10}}{3})F^{LL}_{eK}+\frac{3 C_{10}}{2}M^{LL}_{eK}]\big\} ,\label{amp5}
 %------------------------------------------------------------------------------------------------
\eeq
\beq
   {\cal A}(B_s^0 \to K^0(\rho^0\to)\pi\pi) &=& \frac{G_F} {2}\big\{V_{ub}^*V_{ud}[(C_1+\frac{C_2}{3})F^{LL}_{eK}+C_2 M^{LL}_{eK}]-V_{tb}^*V_{td}[\frac{3 C_8}{2}M^{SP}_{eK}\non
&+&(-\frac{C_3}{3}-C_4+\frac{5 C_9}{3}+C_{10}+\frac{3}{2}(C_7+\frac{C_8}{3}))F^{LL}_{eK}\non
&+&(-C_3+\frac{C_9}{2}+\frac{3 C_{10}}{2})M^{LL}_{eK}-(C_5-\frac{C_7}{2})(M^{LR}_{eK}+M^{LR}_{aK})\non
&-&(\frac{C_3}{3}+C_4-\frac{1}{2}(\frac{C_9}{3}+C_{10}))F^{LL}_{aK})-(C_3-\frac{C_9}{2})M^{LL}_{aK}\non
&-&(\frac{C_5}{3}+C_6-\frac{1}{2}(\frac{C_7}{3}+C_8))F^{SP}_{aK}]\big\} ,\label{amp6}
\eeq
 %------------------------------------------------------------------------------------------------
\item[$\bullet$]  $B_{(s)} \to \pi (\rho \to) \pi \pi$
 \beq
{\cal A}(B^+ \to \pi^+(\rho^0\to)\pi\pi) &=& \frac{G_F} {2}\big\{V_{ub}^*V_{ud}[(\frac{C_1}{3}+C_2)(F^{LL}_{e\rho}+F^{LL}_{a\rho}-F^{LL}_{a\pi})+(C_1+\frac{C_2}{3})F^{LL}_{e\pi}\non
&+&C_1 (M^{LL}_{e\rho}+M^{LL}_{a\rho}-M^{LL}_{a\pi})+C_2 M^{LL}_{e\pi}]-V_{tb}^*V_{td}[\frac{3 C_8}{2} M^{SP}_{e\pi}\non
&+&(\frac{C_3}{3}+C_4+\frac{C_9}{3}+C_{10})(F^{LL}_{e\rho}+F^{LL}_{a\rho}-F^{LL}_{a\pi})\non
&+&(C_3+C_9)(M^{LL}_{e\rho}+M^{LL}_{a\rho}-M^{LL}_{a\pi})+(-C_5+\frac{C_7}{2})M^{LR}_{e\pi}\non
&+&(\frac{C_5}{3}+C_6+\frac{C_7}{3}+C_8)(F^{SP}_{e\rho}+F^{SP}_{a\rho}-F^{SP}_{a\pi})\non
&+&(C_5+C_7)(M^{LR}_{e\rho}+M^{LR}_{a\rho}-M^{LR}_{a\pi})+(-\frac{C_3}{3}-C_4+\frac{5}{3}C_9\non
&+&C_{10}+\frac{3}{2}(C_7+\frac{C_8}{3}))F^{LL}_{e\pi}+(-C_3+\frac{C_9}{2}+\frac{3 C_{10}}{2})M^{LL}_{e\pi}]\big\} ,\label{amp7}
\eeq
 %------------------------------------------------------------------------------------------------
 \beq
 {\cal A}(B^0 \to \pi^-(\rho^+\to)\pi\pi) &=& \frac{G_F} {\sqrt{2}}
 \big\{V_{ub}^*V_{ud}[(C_1+\frac{C_2}{3})F^{LL}_{a\rho}+(\frac{C_1}{3}+C_2)F^{LL}_{e\pi}+C_2 M^{LL}_{a\rho}\non
&+&C_1 M^{LL}_{e\pi}]-V_{tb}^*V_{td}[(\frac{C_3}{3}+C_4+\frac{C_9}{3}+C_{10})F^{LL}_{e\pi}+(C_4\non
&+&C_{10})M^{LL}_{a\rho}+(C_3+\frac{C_4}{3}-C_5-\frac{C_6}{3}-C_7-\frac{C_8}{3}+C_9+\frac{C_{10}}{3})F^{LL}_{a\rho}\non
&+&(C_3+C_9)M^{LL}_{e\pi}+(C_5+C_7)M^{LR}_{e\pi}+(C_5-\frac{C_7}{2})M^{LR}_{a\pi}\non
&+&(\frac{4}{3}(C_3+C_4-\frac{C_9}{2}-\frac{C_{10}}{2})-C_5-\frac{C_6}{3}+\frac{1}{2}(C_7+\frac{C_8}{3}))F^{LL}_{a\pi}\non
&+&(\frac{C_5}{3}+C_6-\frac{1}{2}(\frac{C_7}{3}+C_8))F^{SP}_{a\pi}+(C_6-\frac{C_8}{2})M^{SP}_{a\pi}\non
&+&(C_3+C_4-\frac{C_9}{2}-\frac{C_{10}}{2})M^{LL}_{a\pi}+(C_6+C_8)M^{SP}_{a\rho}]\big\},\label{amp8}
\eeq
 %------------------------------------------------------------------------------------------------
 \beq
 {\cal A}(B^0 \to \pi^+(\rho^-\to)\pi\pi) &=& \frac{G_F} {\sqrt{2}}
 \big\{V_{ub}^*V_{ud}[(\frac{C_1}{3}+C_2)F^{LL}_{e\rho}+(C_1+\frac{C_2}{3})F^{LL}_{a\pi}+C_1 M^{LL}_{e\rho}\non
&+&C_2 M^{LL}_{a\pi}]-V_{tb}^*V_{td}[(\frac{C_3}{3}+C_4+\frac{C_9}{3}+C_{10})F^{LL}_{e\rho}+(C_3\non
&+&C_9)M^{LL}_{e\rho}+(\frac{C_5}{3}+C_6+\frac{C_7}{3}+C_8)F^{SP}_{e\rho}+(C_5+C_7)M^{LR}_{e\rho}\non
&+&(C_6+C_8)M^{SP}_{a\pi}+(\frac{4}{3}(C_3+C_4-\frac{C_9}{2}-\frac{C_{10}}{2})-C_5-\frac{C_6}{3}\non
&+&\frac{1}{2}(C_7+\frac{C_8}{3}))F^{LL}_{a\rho}+(\frac{C_5}{3}+C_6-\frac{1}{2}(\frac{C_7}{3}+C_8))F^{SP}_{a\rho}\non
&+&(C_3+C_4-\frac{C_9}{2}-\frac{C_{10}}{2})M^{LL}_{a\rho}+(C_5-\frac{C_7}{2})M^{LR}_{a\rho}+(C_6\non
&-&\frac{C_8}{2})M^{SP}_{a\rho}+(C_4+C_{10})M^{LL}_{a\pi}+(C_3+\frac{C_4}{3}-C_5-\frac{C_6}{3}-C_7\non
&-&\frac{C_8}{3}+C_9+\frac{C_{10}}{3})F^{LL}_{a\pi}]\big\} ,\label{amp9}
\eeq
 %------------------------------------------------------------------------------------------------
 \beq
 {\cal A}(B_s^0 \to \pi^-(\rho^+\to)\pi\pi) &=& \frac{G_F} {\sqrt{2}}
\big\{V_{ub}^*V_{us}[(C_1+\frac{C_2}{3})F^{LL}_{a\rho}+C_2 M^{LL}_{a\rho}]-V_{tb}^*V_{ts}[(C_6\non
&+&C_8)M^{SP}_{a\rho}+(C_3+\frac{C_4}{3}-C_5-\frac{C_6}{3}-C_7-\frac{C_8}{3}+C_9+\frac{C_{10}}{3})F^{LL}_{a\rho}\non
&+&(C_3+\frac{C_4}{3}-\frac{1}{2}(C_9+\frac{C_{10}}{3})-C_5-\frac{C_6}{3}+\frac{1}{2}(C_7+\frac{C_8}{3}))F^{LL}_{a\pi}\non
&+&(C_4-\frac{C_{10}}{2})M^{LL}_{a\pi}+(C_6-\frac{C_8}{2})M^{SP}_{a\pi}+(C_4+C_{10})M^{LL}_{a\rho}]\big\} ,\label{amp10}
\eeq
%------------------------------------------------------------------------------------------------
\beq
 {\cal A}(B_s^0 \to \pi^+(\rho^-\to)\pi\pi) &=& \frac{G_F} {\sqrt{2}}
 \big\{V_{ub}^*V_{us}[(C_1+\frac{C_2}{3})F^{LL}_{a\pi}+C_2 M^{LL}_{a\pi}]-V_{tb}^*V_{ts}[(C_4\non
&-&\frac{C_{10}}{2})M^{LL}_{a\rho}+(C_3+\frac{C_4}{3}-\frac{1}{2}(C_9+\frac{C_{10}}{3})-C_5-\frac{C_6}{3}\non
&+&\frac{1}{2}(C_7+\frac{C_8}{3}))F^{LL}_{a\rho}+(C_6-\frac{C_8}{2})M^{SP}_{a\rho}+(C_4+C_{10})M^{LL}_{a\pi}\non
&+&(C_6+C_8)M^{SP}_{a\pi}+(C_3+\frac{C_4}{3}-C_5-\frac{C_6}{3}-C_7-\frac{C_8}{3}+C_9\non
&+&\frac{C_{10}}{3})F^{LL}_{a\pi}]\big\} ,\label{amp11}
\eeq
 %------------------------------------------------------------------------------------------------
 \beq
{\cal A}(B^+ \to \pi^0(\rho^+\to)\pi\pi) &=& \frac{G_F} {2}
\big\{V_{ub}^*V_{ud}[(C_1+\frac{C_2}{3})F^{LL}_{e\rho}+(\frac{C_1}{3}+C_2)(-F^{LL}_{a\rho}+F^{LL}_{e\pi}+F^{LL}_{a\pi})\non
&+&C_2 M^{LL}_{e\rho}+C_1(- M^{LL}_{a\rho}+M^{LL}_{e\pi}+M^{LL}_{a\pi})]-V_{tb}^*V_{td}[\frac{3 C_8}{2}M^{SP}_{e\rho}\non
&+&(-\frac{C_3}{3}-C_4-\frac{3}{2}(C_7+\frac{C_8}{3})+\frac{5 C_9}{3}+C_{10})F^{LL}_{e\rho}\non
&+&(-\frac{C_5}{3}-C_6+\frac{1}{2}(\frac{C_7}{3}+C_8))F^{SP}_{e\rho}+(-C_3+\frac{C_9}{2}+\frac{3 C_{10}}{2})M^{LL}_{e\rho}\non
&+&(\frac{C_3}{3}+C_4+\frac{C_9}{3}+C_{10})(-F^{LL}_{a\rho}+F^{LL}_{e\pi}+F^{LL}_{a\pi})\non
&+&(\frac{C_5}{3}+C_6+\frac{C_7}{3}+C_8)(-F^{SP}_{a\rho}+F^{SP}_{a\pi})+(-C_5+\frac{C_7}{2})M^{LR}_{e\rho}\non
&+&(C_3+C_9)(-M^{LL}_{a\rho}+M^{LL}_{e\pi}+M^{LL}_{a\pi})\non
&+&(C_5+C_7)(-M^{LR}_{a\rho}+M^{LR}_{e\pi}+M^{LR}_{a\pi})]\big\},\label{amp12}
\eeq
 %------------------------------------------------------------------------------------------------
 \beq
{\cal A}(B^0 \to \pi^0(\rho^0\to)\pi\pi) &=& -\frac{G_F} {2\sqrt{2}}
 \big\{V_{ub}^*V_{ud}[(C_1+\frac{C_2}{3})(F^{LL}_{e\rho}-F^{LL}_{a\rho}+F^{LL}_{e\pi}-F^{LL}_{a\pi})\non
&+&C_2(M^{LL}_{e\rho}-M^{LL}_{a\rho}+M^{LL}_{e\pi}-M^{LL}_{a\pi})]-V_{tb}^*V_{td}[\frac{3 C_8}{2}(M^{SP}_{e\rho}\non
&+&M^{SP}_{e\pi})+(-\frac{C_3}{3}-C_4-\frac{3}{2}(C_7+\frac{C_8}{3})+\frac{5 C_9}{3}+C_{10})(F^{LL}_{e\rho}\non
&+&F^{LL}_{e\pi})+(-\frac{C_5}{3}-C_6+\frac{1}{2}(\frac{C_7}{3}+C_8))F^{SP}_{e\rho}+(-C_3+\frac{C_9}{2}\non
&+&\frac{3 C_{10}}{2})(M^{LL}_{e\rho}+M^{LL}_{e\pi})+(-C_5+\frac{C_7}{2})(M^{LR}_{e\rho}+M^{LL}_{e\pi})\non
&-&(2C_6+\frac{C_8}{2})(M^{SP}_{a\rho}+M^{SP}_{a\pi})-(\frac{7 C_3}{3}+\frac{5 C_4}{3}-2(C_5+\frac{C_6}{3})\non
&-&\frac{1}{2}(C_7+\frac{C_8}{3}-\frac{2}{3}(C_9-C_{10})))(F^{LL}_{a\rho}+F^{LL}_{a\pi})-(\frac{C_5}{3}+C_6\non
&-&\frac{1}{2}(\frac{C_7}{3}+C_8))(F^{SP}_{a\rho}+F^{SP}_{a\pi})-(C_5-\frac{C_7}{2})(M^{LR}_{a\rho}+M^{LR}_{a\pi})\non
&-&(C_3+2C_4-\frac{C_9}{2}+\frac{C_{10}}{2})(M^{LL}_{a\rho}+M^{LL}_{a\pi})]\big\},\label{amp13}
\eeq
 %------------------------------------------------------------------------------------------------
 \beq
{\cal A}(B_s^0 \to \pi^0(\rho^0\to)\pi\pi) &=& \frac{G_F} {2\sqrt{2}}
\big\{V_{ub}^*V_{us}[(C_1+\frac{C_2}{3})(F^{LL}_{a\rho}+F^{LL}_{a\pi})+C_2(M^{LL}_{a\rho}+M^{LL}_{a\pi})]\non
&-&V_{tb}^*V_{ts}[(2(C_3+\frac{C_4}{3}-C_5-\frac{C_6}{3})-\frac{1}{2}(C_7+\frac{C_8}{3}-C_9\non
&-&\frac{C_{10}}{3}))(F^{LL}_{a\rho}+F^{LL}_{a\pi})+(2C_4+\frac{C_{10}}{2})(M^{LL}_{a\rho}\non
&+&M^{LL}_{a\pi})+(2C_6+\frac{C_8}{2})(M^{SP}_{a\rho}+M^{SP}_{a\pi})]\big\},\label{amp14}
\eeq
 %------------------------------------------------------------------------------------------------
\item[$\bullet$]  $B_{(s)} \to  K (K^* \to) K \pi$
\begin{eqnarray}
{\cal A}(B^+ \to K^+(\bar{K}^{*0} \to)K\pi) &=& \frac{G_F} {\sqrt{2}} \big\{V_{ub}^*V_{ud}[(\frac{C_1}{3}+C_2)F^{LL}_{aK}
+C_1M^{LL}_{aK}]-V_{tb}^*V_{td}[(\frac{C_3}{3}\nonumber\\
&+&C_4-\frac{C_9}{6}-\frac{C_{10}}{2})F^{LL}_{eK}+(C_3-\frac{C_9}{2})M^{LL}_{eK}+(C_5-\frac{C_7}{2})M^{LR}_{eK}\nonumber\\
&+&(\frac{C_3}{3}+C_4+\frac{C_9}{3}+C_{10})F^{LL}_{aK}+(\frac{C_5}{3}+C_6+\frac{C_7}{3}+C_8)F^{SP}_{aK}\nonumber\\
&+&(C_3+C_9)M^{LL}_{aK}+(C_5+C_7)M^{LR}_{aK}]\big\} \;, \label{am1}
\end{eqnarray}
 %------------------------------------------------------------------------------------------------
 \begin{eqnarray}
{\cal A}(B^0 \to K^+(K^{*-} \to)K\pi) &=& \frac{G_F} {\sqrt{2}} \big\{V_{ub}^*V_{ud}[(C_1+\frac{C_2}{3})F^{LL}_{aK}+C_2M^{LL}_{aK}]-V_{tb}^*V_{td}[(C_3\nonumber\\
&+&\frac{C_4}{3}-\frac{C_9}{2}-\frac{C_{10}}{6}-C_5-\frac{C_6}{3}+\frac{C_7}{2}+\frac{C_8}{6})F^{LL}_{aK^*}\nonumber\\
&+&(C_4-\frac{C_{10}}{2})M^{LL}_{aK^*}+(C_6-\frac{C_8}{2})M^{SP}_{aK^*}+(C_3+\frac{C_4}{3}\nonumber\\
&+&C_9+\frac{C_{10}}{3}-C_5-\frac{C_6}{3}-C_7-\frac{C_8}{3})F^{LL}_{aK}\nonumber\\
&+&(C_4+C_{10})M^{LL}_{aK}+(C_6+C_8)M^{SP}_{aK}]\big\} \;,
\label{am2}
\end{eqnarray}
%------------------------------------------------------------------------------------------------
\begin{eqnarray}
{\cal A}(B^0 \to K^-(K^{*+} \to)K\pi) &=& \frac{G_F} {\sqrt{2}} \big\{V_{ub}^*V_{ud}[(C_1+\frac{C_2}{3})F^{LL}_{aK^*}+C_2 M^{LL}_{aK^*}]-V_{tb}^*V_{td}[(C_3+\frac{C_4}{3}\nonumber\\
&+&C_9+\frac{C_{10}}{3}-C_5-\frac{C_6}{3}-C_7-\frac{C_8}{3})F^{LL}_{aK^*}+(C_4+C_{10})M^{LL}_{aK^*}\nonumber\\
&+&(C_6+C_8)M^{SP}_{aK^*}+(C_3+\frac{C_4}{3}-\frac{C_9}{2}-\frac{C_{10}}{6}-C_5-\frac{C_6}{3}+\frac{C_7}{2}\nonumber\\
&+&\frac{C_8}{6})F^{LL}_{aK}+(C_4-\frac{C_{10}}{2})M^{LL}_{aK}+(C_6-\frac{C_8}{2})M^{SP}_{aK}]\big\} \;,\nonumber\\
\label{am3}
\end{eqnarray}
%------------------------------------------------------------------------------------------------
\begin{eqnarray}
 {\cal A}(B_s^0 \to K^+(K^{*-}\to)K\pi) &=&  \frac{G_F} {\sqrt{2}} \big\{V_{ub}^*V_{us}[(\frac{C_1}{3}
 +C_2)F^{LL}_{eK^*}+C_1 M^{LL}_{eK^*}+(C_1+\frac{C_2}{3})F^{LL}_{aK}\nonumber\\
&+&C_2M^{LL}_{aK}]-V_{tb}^*V_{ts}[(\frac{C_3}{3}+C_4+\frac{C_9}{3}+C_{10})F^{LL}_{eK^*}+(\frac{C_5}{3}\nonumber\\
&+&C_6+\frac{C_7}{3}+C_8)F^{SP}_{eK^*}+(C_3+C_9)M^{LL}_{eK^*}+(C_5+C_7)M^{LR}_{eK^*}\nonumber\\
&+&(\frac{4}{3}(C_3+C_4-\frac{C_9}{2}-\frac{C_{10}}{2})-C_5-\frac{C_6}{3}+\frac{C_7}{2}\nonumber\\
&+&\frac{C_8}{6})F^{LL}_{aK^*}+(\frac{C_5}{3}+C_6-\frac{C_7}{6}-\frac{C_8}{2})F^{SP}_{aK^*}+(C_3+C_4\nonumber\\
&-&\frac{C_9}{2}-\frac{C_{10}}{2})M^{LL}_{aK^*}+(C_5-\frac{C_7}{2})M^{LR}_{aK^*}+(C_6-\frac{C_8}{2})M^{SP}_{aK^*}\nonumber\\
&+&(C_3+\frac{C_4}{3}+C_9+\frac{C_{10}}{3}-C_5-\frac{C_6}{3}\nonumber\\
&-&C_7-\frac{C_8}{3})F^{LL}_{aK}+(C_4+C_{10})M^{LL}_{aK}+(C_6+C_8)M^{SP}_{aK}]\big\} \;,\label{am4}
\end{eqnarray}
%------------------------------------------------------------------------------------------------
\begin{eqnarray}
 {\cal A}(B_s^0 \to K^-(K^{*+}\to)K \pi) &=&  \frac{G_F} {\sqrt{2}} \big\{V_{ub}^*V_{us}[(C_1+\frac{C_2}{3})F^{LL}_{aK^*}+C_2 M^{LL}_{aK^*}+(\frac{C_1}{3}+C_2)F^{LL}_{eK}\nonumber\\
&+&C_1M^{LL}_{eK}]-V_{tb}^*V_{ts}[(C_3+\frac{C_4}{3}+C_9+\frac{C_{10}}{3}-C_5-\frac{C_6}{3}-C_7\nonumber\\
&-&\frac{C_8}{3})F^{LL}_{aK^*}+(C_4+C_{10})M^{LL}_{aK^*}+(C_6+C_8)M^{SP}_{aK^*}+(\frac{C_3}{3}\nonumber\\
&+&C_4+\frac{C_9}{3}+C_{10})F^{LL}_{eK}+(C_3+C_9)M^{LL}_{eK}+(C_5+C_7)M^{LR}_{eK}\nonumber\\
&+&(\frac{4}{3}(C_3+C_4-\frac{C_9}{2}-\frac{C_{10}}{2})-C_5-\frac{C_6}{3}+\frac{C_7}{2}+\frac{C_8}{6})F^{LL}_{aK}\nonumber\\
&+&(\frac{C_5}{3}+C_6-\frac{C_7}{6}-\frac{C_8}{2})F^{SP}_{aK}+(C_3+C_4-\frac{C_9}{2}\nonumber\\
&-&\frac{C_{10}}{2})M^{LL}_{aK}+(C_5-\frac{C_7}{2})M^{LR}_{aK}+(C_6-\frac{C_8}{2})M^{SP}_{aK}
]\big\} \;,\label{am5}
\end{eqnarray}
%------------------------------------------------------------------------------------------------
\begin{eqnarray}
 {\cal A}(B^+ \to \bar{K}^0(K^{*+}\to)K \pi) &=& \frac{G_F} {\sqrt{2}}\big\{V_{ub}^*V_{ud}[(\frac{C_1}{3}+C_2)F^{LL}_{aK^*}+C_1 M^{LL}_{aK^*}]-V_{tb}^*V_{td}[(\frac{C_3}{3}+C_4\nonumber\\
&-&\frac{C_9}{6}-\frac{C_{10}}{2})F^{LL}_{eK^*}+(\frac{C_5}{3}+C_6-\frac{C_7}{6}-\frac{C_8}{2})F^{SP}_{eK^*}+(C_3\nonumber\\
&-&\frac{C_9}{2})M^{LL}_{eK^*}+(C_5-\frac{C_7}{2})M^{LR}_{eK^*}+(\frac{C_3}{3}+C_4+\frac{C_9}{3}+C_{10})F^{LL}_{aK^*}\nonumber\\
&+&(\frac{C_5}{3}+C_6+\frac{C_7}{3}+C_8)F^{SP}_{aK^*}+(C_3+C_9)M^{LL}_{aK^*}\nonumber\\
&+&(C_5+C_7)M^{LR}_{aK^*}]\big\} \;,\label{am6}
\end{eqnarray}
%------------------------------------------------------------------------------------------------
\begin{eqnarray}
{\cal A}(B^0 \to K^0(\bar{K}^{*0}\to)K \pi) &=& -\frac{G_F} {\sqrt{2}}\big\{V_{tb}^*V_{td}[(C_3+\frac{C_4}{3}-\frac{C_9}{2}-\frac{C_{10}}{6}-C_5-\frac{C_6}{3}+\frac{C_7}{2}\nonumber\\
&+&\frac{C_8}{6})F^{LL}_{aK^*}+(C_4-\frac{C_{10}}{2})M^{LL}_{aK^*}+(C_6-\frac{C_8}{2})(M^{SP}_{aK^*}+M^{SP}_{aK})\nonumber\\
&+&(\frac{C_3}{3}+C_4-\frac{C_9}{6}-\frac{C_{10}}{2})F^{LL}_{eK}+(C_3-\frac{C_9}{2})M^{LL}_{eK}+(C_5\nonumber\\
&-&\frac{C_7}{2})(M^{LR}_{eK}+M^{LR}_{aK})+(\frac{C_5}{3}+C_6-\frac{C_7}{6}-\frac{C_8}{2})F^{SP}_{aK}+(\frac{4}{3}(C_3\nonumber\\
&+&C_4-\frac{C_9}{2}-\frac{C_{10}}{2})-C_5-\frac{C_6}{3}+\frac{C_7}{2}+\frac{C_8}{6})F^{LL}_{aK}+(C_3+C_4\nonumber\\
&-&\frac{C_9}{2}-\frac{C_{10}}{2})M^{LL}_{aK}]\big\} \;,\nonumber\\
\label{am7}
 \end{eqnarray}
 %------------------------------------------------------------------------------------------------
 \begin{eqnarray}
{\cal A}(B^0 \to \bar{K}^0(K^{*0}\to)K \pi) &=& -\frac{G_F} {\sqrt{2}}\big\{V_{tb}^*V_{td}[(\frac{C_3}{3}+C_4-\frac{C_9}{6}-\frac{C_{10}}{2})F^{LL}_{eK^*}+(\frac{C_5}{3}+C_6-\frac{C_7}{6}\nonumber\\
&-&\frac{C_8}{2})(F^{SP}_{eK^*}+F^{SP}_{aK^*})+(C_3-\frac{C_9}{2})M^{LL}_{eK^*}+(\frac{4}{3}(C_3+C_4-\frac{C_9}{2}\nonumber\\
&-&\frac{C_{10}}{2})-C_5-\frac{C_6}{3}+\frac{C_7}{2}+\frac{C_8}{6})F^{LL}_{aK^*}+(C_3+C_4-\frac{C_9}{2}\nonumber\\
&-&\frac{C_{10}}{2})M^{LL}_{aK^*}+(C_5-\frac{C_7}{2})(M^{LR}_{eK^*}+M^{LR}_{aK^*})+(C_6\nonumber\\
&-&\frac{C_8}{2})(M^{SP}_{aK^*}+M^{SP}_{aK})+(C_3+\frac{C_4}{3}-\frac{C_9}{2}-\frac{C_{10}}{6}-C_5\nonumber\\
&-&\frac{C_6}{3}+\frac{C_7}{2}+\frac{C_8}{6})F^{LL}_{aK}+(C_4-\frac{C_{10}}{2})M^{LL}_{aK}]\big\} \;,\nonumber\\
\label{am8}
 \end{eqnarray}
 %------------------------------------------------------------------------------------------------
 \begin{eqnarray}
{\cal A}(B_s^0 \to K^0(\bar{K}^{*0}\to)K \pi) &=& -\frac{G_F} {\sqrt{2}}\big\{V_{tb}^*V_{ts}[(\frac{C_3}{3}+C_4-\frac{C_9}{6}-\frac{C_{10}}{2})F^{LL}_{eK^*}+(\frac{C_5}{3}+C_6-\frac{C_7}{6}\nonumber\\
&-&\frac{C_8}{2})(F^{SP}_{eK^*}+F^{SP}_{aK^*})+(C_3-\frac{C_9}{2})M^{LL}_{eK^*}+(\frac{4}{3}(C_3+C_4-\frac{C_9}{2}\nonumber\\
&-&\frac{C_{10}}{2})-C_5-\frac{C_6}{3}+\frac{C_7}{2}+\frac{C_8}{6})F^{LL}_{aK^*}+(C_3+C_4-\frac{C_9}{2}\nonumber\\
&-&\frac{C_{10}}{2})M^{LL}_{aK^*}+(C_5-\frac{C_7}{2})(M^{LR}_{eK^*}+M^{LR}_{aK^*})+(C_6-\frac{C_8}{2})(M^{SP}_{aK^*}\nonumber\\
&+&M^{SP}_{aK})+(C_3+\frac{C_4}{3}-\frac{C_9}{2}-\frac{C_{10}}{6}-C_5-\frac{C_6}{3}+\frac{C_7}{2}+\frac{C_8}{6})F^{LL}_{aK}\nonumber\\
&+&(C_4-\frac{C_{10}}{2})M^{LL}_{aK}]\big\}  \;,\nonumber\\
\label{am9}
\end{eqnarray}
 %------------------------------------------------------------------------------------------------
\begin{eqnarray}
{\cal A}(B_s^0 \to \bar{K}^0(K^{*0}\to)K \pi) &=& -\frac{G_F} {\sqrt{2}}\big\{V_{tb}^*V_{ts}[(C_3+\frac{C_4}{3}-\frac{C_9}{2}-\frac{C_{10}}{6}-C_5-\frac{C_6}{3}+\frac{C_7}{2}\nonumber\\
&+&\frac{C_8}{6})F^{LL}_{aK^*}+(C_4-\frac{C_{10}}{2})M^{LL}_{aK^*}+(C_6-\frac{C_8}{2})(M^{SP}_{aK^*}+M^{SP}_{aK})\nonumber\\
&+&(\frac{C_3}{3}+C_4-\frac{C_9}{6}-\frac{C_{10}}{2})F^{LL}_{eK}+(C_3-\frac{C_9}{2})M^{LL}_{eK}+(C_5-\frac{C_7}{2})(M^{LR}_{eK}\nonumber\\
&+&M^{LR}_{aK})+(\frac{C_5}{3}+C_6-\frac{C_7}{6}-\frac{C_8}{2})F^{SP}_{aK}+(\frac{4}{3}(C_3+C_4-\frac{C_9}{2}\nonumber\\
&-&\frac{C_{10}}{2})-C_5-\frac{C_6}{3}+\frac{C_7}{2}+\frac{C_8}{6})F^{LL}_{aK}+(C_3+C_4-\frac{C_9}{2}\nonumber\\
&-&\frac{C_{10}}{2})M^{LL}_{aK}]\big\} \;,\nonumber\\
\label{am10}
\end{eqnarray}
 %------------------------------------------------------------------------------------------------
 \item[$\bullet$]  $B_{(s)} \to  \pi (K^* \to) K \pi$
 \begin{eqnarray}
{\cal A}(B^+ \to \pi^+(K^{*0}\to)K \pi) &=& \frac{G_F} {\sqrt{2}}\big\{V_{ub}^*V_{us}[(\frac{C_1}{3}+C_2)F^{LL}_{a\pi}+C_1M^{LL}_{a\pi}]-V_{tb}^*V_{ts}[(\frac{C_3}{3}+C_4\nonumber\\
&-&\frac{C_9}{6}-\frac{C_{10}}{2})F^{LL}_{e\pi}+(C_3-\frac{C_9}{2})M^{LL}_{e\pi}+(C_5-\frac{C_7}{2})M^{LR}_{e\pi}+(\frac{C_3}{3}\nonumber\\
&+&C_4+\frac{C_9}{3}+C_{10})F^{LL}_{a\pi}+(\frac{C_5}{3}+C_6+\frac{C_7}{3}+C_8)F^{SP}_{a\pi}\nonumber\\
&+&(C_3+C_9)M^{LL}_{a\pi}+(C_5+C_7)M^{LR}_{a\pi}]\big\} \;,\nonumber\\
 \label{am11}
\end{eqnarray}
 %------------------------------------------------------------------------------------------------
 \begin{eqnarray}
 {\cal A}(B^0 \to \pi^-(K^{*+}\to)K \pi) &=& \frac{G_F} {\sqrt{2}}\big\{V_{ub}^*V_{us}[(\frac{C_1}{3}+C_2)F^{LL}_{e\pi}+C_1M^{LL}_{e\pi}]-V_{tb}^*V_{ts}[(\frac{C_3}{3}+C_4\nonumber\\
 &+&\frac{C_9}{3}+C_{10})F^{LL}_{e\pi}+(C_3+C_9)M^{LL}_{e\pi}+(C_5+C_7)M^{LR}_{e\pi}+(\frac{C_3}{3}\nonumber\\
 &+&C_4-\frac{C_9}{6}-\frac{C_{10}}{2})F^{LL}_{a\pi}+(\frac{C_5}{3}+C_6-\frac{C_7}{6}-\frac{C_8}{2})F^{SP}_{a\pi}+(C_3\nonumber\\
 &-&\frac{C_9}{2})M^{LL}_{a\pi}+(C_5-\frac{C_7}{2})M^{LR}_{a\pi}]\big\} \;,\label{am12}
\end{eqnarray}
 %------------------------------------------------------------------------------------------------
 \begin{eqnarray}
 {\cal A}(B_s^0 \to \pi^+(K^{*-}\to)K\pi) &=& \frac{G_F} {\sqrt{2}}\big\{V_{ub}^*V_{ud}[(\frac{C_1}{3}+C_2)F^{LL}_{eK^*}+C_1M^{LL}_{eK^*}]-V_{tb}^*V_{td}[(\frac{C_3}{3}+C_4\nonumber\\
 &+&\frac{C_9}{3}+C_{10})F^{LL}_{eK^*}+(\frac{C_5}{3}+C_6+\frac{C_7}{3}+C_8)F^{SP}_{eK^*}+(C_3+C_9)M^{LL}_{eK^*}\nonumber\\
 &+&(C_5+C_7)M^{LR}_{eK^*}+(\frac{C_3}{3}+C_4-\frac{C_9}{6}-\frac{C_{10}}{2})F^{LL}_{aK^*}+(\frac{C_5}{3}+C_6\nonumber\\
 &-&\frac{C_7}{6}-\frac{C_8}{2})F^{SP}_{aK^*}+(C_3-\frac{C_9}{2})M^{LL}_{aK^*}+(C_5-\frac{C_7}{2})M^{LR}_{aK^*}]\big\}  \;,\label{am13}
\end{eqnarray}
 %------------------------------------------------------------------------------------------------
\begin{eqnarray}
{\cal A}(B^+ \to \pi^0(K^{*+}\to)K \pi) &=& \frac{G_F} {2}
\big\{V_{ub}^*V_{us}[(C_1+\frac{C_2}{3})F^{LL}_{eK^*}+(\frac{C_1}{3}+C_2)(F^{LL}_{e\pi}+F^{LL}_{a\pi})\nonumber\\
&+&C_1(M^{LL}_{e\pi}+M^{LL}_{a\pi})+C_2M^{LL}_{eK^*}]-V_{tb}^*V_{ts}[(\frac{3C_9}{2}+\frac{C_{10}}{2}-\frac{3C_7}{2}\nonumber\\
&-&\frac{C_8}{2})F^{LL}_{eK^*}+\frac{3C_{10}}{2}M^{LL}_{eK^*}+\frac{3C_8}{2}M^{SP}_{eK^*}+(\frac{C_3}{3}+C_4+\frac{C_9}{3}\nonumber\\
&+&C_{10})(F^{LL}_{e\pi}+F^{LL}_{a\pi})+(\frac{C_5}{3}+C_6+\frac{C_7}{3}+C_8)F^{SP}_{a\pi}+(C_3\nonumber\\
&+&C_9)(M^{LL}_{e\pi}+M^{LL}_{a\pi})+(C_5+C_7)(M^{LR}_{e\pi}+M^{LR}_{a\pi})]\big\} \;,\label{am14}
\end{eqnarray}
 %------------------------------------------------------------------------------------------------
\begin{eqnarray}
 {\cal A}(B^0 \to \pi^0(K^{*0}\to)K \pi) &=& \frac{G_F} {2}
\big\{V_{ub}^*V_{us}[(C_1+\frac{C_2}{3})F^{LL}_{eK^*}+C_2M^{LL}_{eK^*}]-V_{tb}^*V_{ts}[(\frac{3C_9}{2}+\frac{C_{10}}{2}\nonumber\\
&-&\frac{3C_7}{2}-\frac{C_8}{2})F^{LL}_{eK^*}+\frac{3C_{10}}{2}M^{LL}_{eK^*}+\frac{3C_8}{2}M^{SP}_{eK^*}-(\frac{C_3}{3}+C_4-\frac{C_9}{6}\nonumber\\
&-&\frac{C_{10}}{2})(F^{LL}_{e\pi}+F^{LL}_{a\pi})-(\frac{C_5}{3}+C_6-\frac{C_7}{6}-\frac{C_8}{2})F^{SP}_{a\pi}-(C_3\nonumber\\
&-&\frac{C_9}{2})(M^{LL}_{e\pi}+M^{LL}_{a\pi})-(C_5-\frac{C_7}{2})(M^{LR}_{e\pi}+M^{LR}_{a\pi})]\big\} \;,\nonumber\\
\label{am15}
\end{eqnarray}
 %------------------------------------------------------------------------------------------------
 \begin{eqnarray}
{\cal A}(B_s^0 \to \pi^0(\bar{K}^{*0}\to)K \pi) &=& \frac{G_F} {2}
\big\{V_{ub}^*V_{ud}[(C_1+\frac{C_2}{3})F^{LL}_{eK^*}+C_2M^{LL}_{eK^*}]-V_{tb}^*V_{td}[(-\frac{C_3}{3}-C_4\nonumber\\
&+&\frac{5C_9}{3}+C_{10}-\frac{3C_7}{2}-\frac{C_8}{2})F^{LL}_{eK^*}-(\frac{C_5}{3}+C_6-\frac{C_7}{6}-\frac{C_8}{2})F^{SP}_{eK^*}\nonumber\\
&+&(-C_3+\frac{C_9}{2}+\frac{3C_{10}}{2})M^{LL}_{eK^*}-(C_5-\frac{C_7}{2})M^{LR}_{eK^*}+\frac{3C_8}{2}M^{SP}_{eK^*}\nonumber\\
&-&(\frac{C_3}{3}+C_4-\frac{C_9}{6}-\frac{C_{10}}{2})F^{LL}_{aK^*}+(\frac{C_5}{3}+C_6-\frac{C_7}{6}-\frac{C_8}{2})F^{SP}_{aK^*}\nonumber\\
&-&(C_3-\frac{C_9}{2})M^{LL}_{aK^*}-(C_5-\frac{C_7}{2})M^{LR}_{aK^*}]\big\} \;,\label{am16}
\end{eqnarray}
%------------------------------------------------------------------------------------------------
\item[$\bullet$]  $B_{(s)} \to  K (\phi \to) K K$
%----------------------------------
\begin{eqnarray}
{\cal A}(B^+ \to K^+ (\phi\to)K K)& =& \frac{G_F}{\sqrt{2}}\big\{V^{*}_{ub}V_{us}[(\frac{C_1}{3} +C_2 )F_{aK}^{LL}+C_1M_{aK}^{LL}]-\frac{G_F}{\sqrt{2}}V^{*}_{tb}V_{ts}[(\frac{4}{3}(C_3+C_4)\non
&+&C_5+\frac{C_6}{3}-\frac{C_{7}}{2}-\frac{C_{8}}{6}-\frac{2}{3}(C_9+C_{10}))F_{eK}^{LL}+(C_3+C_4-\frac{1}{2}(C_{9}\non
  &+&C_{10}))M_{eK}^{LL}+(C_5-\frac{C_{7}}{2}) M_{eK}^{LR}+(C_6-\frac{C_{8}}{2}) M_{eK}^{SP}+(\frac{C_{3}}{3}+C_4\non
  &+&\frac{C_{9}}{3}+C_{10})F_{aK}^{LL}+(\frac{C_{5}}{3}+C_6+\frac{C_{7}}{3}+C_{8})F_{aK}^{SP}+(C_3+C_{9}) M_{aK}^{LL}\non
  &+&(C_5+C_{7}) M_{aK}^{LR} ]\big\} \;,
\end{eqnarray}
%----------------------------------
\begin{eqnarray}
 {\cal A}(B^0 \to K^0 (\phi\to)K K)& =&-\frac{G_F}{\sqrt{2}}\big\{V^{*}_{tb}V_{ts}[(\frac{4}{3}C_3+\frac{4}{3}C_4+C_5+\frac{C_6}{3}-\frac{C_{7}}{2}-\frac{C_{8}}{6}-\frac{2}{3}C_9\non
 &-&\frac{2}{3}C_{10})F_{eK}^{LL}+(C_3+C_4-\frac{C_{9}}{2}-\frac{C_{10}}{2})M_{eK}^{LL}+(C_5-\frac{C_{7}}{2}) M_{eK}^{LR}\non
 &+&(C_6-\frac{C_{8}}{2})M_{eK}^{SP}+(\frac{C_3}{3}+C_4-\frac{C_{9}}{6}-\frac{C_{10}}{2})F_{aK}^{LL}+(\frac{C_5}{3}+C_6\non
 &-&\frac{C_{7}}{6}-\frac{C_{8}}{2})F_{aK}^{SP}+(C_3-\frac{C_{9}}{2}) M_{aK}^{LL}+(C_5-\frac{C_{7}}{2}) M_{aK}^{LR}]\big\} \;,
\end{eqnarray}
%----------------------------------
\begin{eqnarray}
A( B_{s}^0\to    \bar K^{0} (\phi\to)K K) &=& - \frac{G_F}{\sqrt{2}}
\big\{V^{*}_{tb}V_{td} [(C_3+\frac{C_4}{3}+C_5+\frac{C_6}{3}-\frac{C_{7}}{2}-\frac{C_{8}}{6}-\frac{C_9}{2}-\frac{C_{10}}{6})F_{e K}^{LL}\non
 &+&(\frac{C_3}{3}+C_4-\frac{C_{9}}{6}-\frac{C_{10}}{2})(F_{e\phi}^{LL}+F_{a\phi}^{LL})+(\frac{C_5}{3}+C_6
 -\frac{C_7}{6}-\frac{C_8}{2})(F_{e\phi}^{SP}\non
 &+&F_{a\phi}^{SP})+(C_{3}-\frac{C_9}{2})(M_{e\phi}^{LL}+M_{a\phi}^{LL})
 +(C_{5}- \frac{C_7}{2})(M_{e\phi}^{LR}+M_{a\phi}^{LR})\non
 &+&(C_{4}-\frac{C_{10}}{2})M_{eK}^{LL}+(C_{6}-\frac{C_8}{2})M_{eK}^{SP}]\big\} \;,
\end{eqnarray}
%-------------------------------------------------
\item[$\bullet$]  $B_{(s)} \to  \pi (\phi \to) K K$
%----------------------------------
\begin{eqnarray}
  {\cal A}(B^+ \to \pi^+ (\phi\to)K K)& =& -\frac{G_F}{\sqrt{2}}\big\{V^{*}_{tb}V_{td}[(C_3+\frac{C_4}{3}+C_5+\frac{C_6}{3}-\frac{C_{7}}{2}-\frac{C_{8}}{6}-\frac{C_{9}}{2}  \nonumber \\
  &-&\frac{C_{10}}{6})F_{e\pi}^{LL}+(C_4-\frac{C_{10}}{2}) M_{e\pi}^{LL}+(C_6-\frac{C_{8}}{2}) M_{e\pi}^{SP}]\big\}\;,
\end{eqnarray}
%----------------------------------
\begin{eqnarray}
 \sqrt{2}{\cal A}(B^0 \to \pi^0 (\phi\to)K K)& =&-\frac{G_F}{\sqrt{2}}\big\{V^{*}_{tb}V_{td}[(-C_3-\frac{C_4}{3}-C_5-\frac{C_6}{3}+\frac{C_{7}}{2}+\frac{C_{8}}{6}+\frac{C_{9}}{2}\nonumber \\
 &+&\frac{C_{10}}{6})F_{e\pi}^{LL}+(-C_4+\frac{C_{10}}{2})M_{e\pi}^{LL}+(-C_6+\frac{C_{8}}{2})M_{e\pi}^{SP}]\big\} \;,
\end{eqnarray}
%----------------------------------
\begin{eqnarray}
\sqrt{2}A(B_{s}^0\to\pi^{0}(\phi\to)K K) &=& \frac{G_F}{\sqrt{2}}
\big\{V^{*}_{ub}V_{us} [(C_1+\frac{C_2}{3})F^{LL}_{e\phi}+C_2M_{e\phi}^{LL}]- \frac{G_F}{\sqrt{2}}V^{*}_{tb}V_{ts}[(-\frac{3}{2}C_{7}\non
&-&\frac{C_{8}}{2}+\frac{3}{2}C_9+\frac{C_{10}}{2})F^{LL}_{e\phi}+\frac{3}{2}C_{8}M_{e\phi}^{SP}+\frac{3}{2}C_{10}M_{e\phi}^{LL}]\big\} \;.
\end{eqnarray}

\end{enumerate}

The explicit PQCD factorization formulas for the functions $F$ and $M$ appearing in the above decay amplitudes
are given by
%Performing the standard PQCD calculations, one gets the following expressions of the relevant amplitudes $F_{e(a)}$ and $M_{e(a)}$:
%\begin{enumerate}
%\item[$\bullet$]  $B_{(s)} \to PV$ decay modes (The meson $P$ is moving along the direction of $v$ and resonance $V$ is along $n$):
\beq
F^{LL}_{eV}&=&8 \pi C_F f_P m_B^4 \int_0^1 dx_B dz \int_0^\infty b_B\; db_B\; b\; db\; \phi_B(x_B,b_B)\non
&\times&\bigg\{\big[
((f_{-} g_{-}-f_{+} g_{+} (1+f_{+} z))\sqrt{\eta }\phi_0(z)-(g_{-}+g_{+}(1-2 f_{+} z)) \eta \phi_s(z)\non
&+&f_{-} f_{+} (g_{-}+g_{+} (-1+2 f_{+} z)) \phi_t(z))/\sqrt{\eta}
\big] E_e(t_a^V)\; h^V_a(\alpha^V_e,\beta^V_a,b_B,b)S_t(z)\non
&+&\big[
f_{+} (f_{-} (g_{-}-g_{+})-g_{-} x_B) \phi_0(z)
-2 ((f_{-}-x_B) g_{-}+f_{+} g_{+}) \sqrt{\eta }\phi_s(z)
 \big]\non
&\times&\; E_e(t_b^V)\; h^V_b(\alpha^V_e,\beta^V_b,b_B,b)S_t(|x_B-f_{-}|)\bigg\},\label{eq:fll01}
\eeq
\beq
F^{LR}_{eV}&=&-F^{LL}_{eV}, \label{eq:flr01}
\eeq
\beq
F^{SP}_{eV}&=&16\pi C_F f_P m_B^4 r_{03}\int_0^1 dx_B dz \int_0^\infty b_B\; db_B\; b\; db\; \phi_B(x_B,b_B)\non
&\times&\bigg\{\big[
((f_{+}+f_{-} (-1+2 f_{+} z)) \sqrt{\eta } \phi_0(z)+(2+f_{+} z) \eta \phi_s(z)-f_{-} f_{+}^2 z \phi_t(z))/\sqrt{\eta}
\big] \non
&\times& E_e(t_a^V)\; h^V_a(\alpha^V_e,\beta^V_a,b_B,b)S_t(z)\non
&+&\big[
f_{+} x_B \phi_0(z)+2 (f_{-}+f_{+}-x_B) \sqrt{\eta }\phi_s(z)
\big] E_e(t_b^V)\; h^V_b(\alpha^V_e,\beta^V_b,b_B,b)S_t(|x_B-f_{-}|)\bigg\},\label{eq:fsp01}
\eeq
\beq
M^{LL}_{eV}&=&32\pi C_F m^4_B/\sqrt{6}/\sqrt{\eta} \int_0^1 dx_B dz dx_3
\int_0^\infty b_B db_B\; b_3 db_3\; \phi_B(x_B,b_B)\phi^A_P(x_3)\non
&\times&\bigg\{\big[
(g_{-}+g_{+}) (f_{+} (g_{+} (-1+x_3)+x_B)+f_{-} (g_{-}+f_{+} z)) \sqrt{\eta } \phi_0(z)+(-g_{-} (g_{+} (x_3-2)\non
&+&x_B)+f_{+} g_{+} z) \eta \phi_s(z)-f_{-} f_{+} (g_{-} (g_{+} x_3+x_B)+f_{+} g_{+} z) \phi_t(z)
\big]\; E_n(t_c^V)\; h^V_c(\alpha^V_e,\beta^V_c,b_B,b_3)\non
&-&\big[
(f_{-} g_{-}-f_{+} g_{+}) (g_{+} x_3-x_B+f_{+} z) \sqrt{\eta } \phi_0(z)+(g_{-} g_{+} x_3-g_{-} x_B+f_{+} g_{+} z) \eta \phi_s(z)\non
&+&f_{-} f_{+} (g_{-} (-g_{+} x_3+x_B)+f_{+} g_{+} z) \phi_t(z)\big] \; E_n(t_d^V)\; h^V_d(\alpha^V_e,\beta^V_d,b_B,b_3)
\bigg\},\label{eq:mll01}
\eeq
\beq
M^{LR}_{eV}&=&-32\pi C_F m^4_B r_{03} /\sqrt{6}/\sqrt{\eta} \int_0^1 dx_B dz dx_3 \int_0^\infty b_B db_B\; b_3 db_3\; \phi_B(x_B,b_B)\non
&\times&\bigg\{\big[
\sqrt{\eta } \phi_0(z)  (f_{-} (g_{-}+f_{+} z) (\phi_P^P(x_3)-\phi_P^T(x_3))+f_{+} (g_{+} (x_3-1)+x_B) (\phi_P^P(x_3)+\phi_P^T(x_3)))\non
&-&(\eta\phi_s(z)+f_{+}f_{-}\phi_t(z))(f_{+} z+g_{-}) (\phi_P^P(x_3)-\phi_P^T(x_3))
+(\eta\phi_s(z)-f_{+}f_{-}\phi_t(z))(x_B\non
&+&g_{+} (x_3-1)) (\phi_P^P(x_3)+\phi_P^T(x_3))\big]E_n(t_c^V)\; h^V_c(\alpha^V_e,\beta^V_c,b_B,b_3)\non
&+&\big[
f_{+} \sqrt{\eta } \phi_0(z)  ((g_{+} x_3-x_B) (\phi_P^P(x_3)-\phi_P^T(x_3))-f_{-} z (\phi_P^P(x_3)+\phi_P^T(x_3)))\non
&+&(g_{+} x_3-x_B)(\phi_P^P(x_3)-\phi_P^T(x_3))(\eta\phi_s(z)-f_{+}f_{-}\phi_t(z))\non
&+&f_{+} z(\phi_P^P(x_3)+\phi_P^T(x_3))(\eta\phi_s(z)+f_{+}f_{-}\phi_t(z))\big] E_n(t_d^V)\; h^V_d(\alpha^V_e,\beta^V_d,b_B,b_3)
\bigg\},\label{eq:mlr01}
\eeq
\beq
M^{SP}_{eV}&=&-32\pi C_F m^4_B /\sqrt{6}/\sqrt{\eta} \int_0^1 dx_B dz dx_3
\int_0^\infty b_B db_B\; b_3 db_3\; \phi_B(x_B,b_B)\phi^A_P(x_3)\non
&\times&\bigg\{\big[
(f_{-} g_{-}-f_{+} g_{+}) (g_{-}+g_{+}(1- x_3)
-x_B+f_{+} z) \sqrt{\eta } \phi_0(z)+(-g_{-} (g_{+} (-2\non
&+&x_3)+x_B)+f_{+} g_{+} z) \eta \phi_s(z)+f_{-} f_{+} (g_{-} (g_{+} x_3+x_B)+f_{+} g_{+} z) \phi_t(z)
\big]\non
&\times& E_n(t_c^V)\; h^V_c(\alpha^V_e,\beta^V_c,b_B,b_3)\non
&-&\big[-f_{+} (g_{-}+g_{+}) (g_{+} x_3-x_B-f_{-} z) \sqrt{\eta } \phi_0(z)
+(g_{-} g_{+} x_3-g_{-} x_B+f_{+} g_{+} z)\non
&\times&\eta \phi_s(z)-f_{-} f_{+} (g_{-} (x_B-g_{+} x_3)+f_{+} g_{+} z) \phi_t(z)
\big] E_n(t_d^V)\; h^V_d(\alpha^V_e,\beta^V_d,b_B,b_3)
\bigg\},\label{eq:msp01}
\eeq
\beq
F^{LL}_{aV}&=&-8\pi C_F m^4_B f_B/\sqrt{6} \int_0^1 dz dx_3 \int_0^\infty b db\; b_3 db_3\;\non
&\times&\bigg\{\big[
((f_{-}^2 g_{-}+f_{-} g_{-} g_{+}+f_{+} g_{+} (f_{+} (z-1)-g_{-})) \sqrt{\eta } \phi_0(z)  \phi_P^A(x_3)
+2 r_{03} (f_{-}+f_{+}+g_{-}+g_{+}\non
&-&f_{+} z) \eta\phi_s(z) \phi_P^P(x_3)
+2 f_{-} f_{+} r_{03} (f_{-}-g_{-}+g_{+}+f_{+} (z-1)) \phi_t(z)\phi_P^P(x_3))/\sqrt{\eta}
\big]E_a(t_e^V)\; h^V_e(\alpha^V_a,\beta^V_e,b,b_3)S_t(z)\non
&-&\big[
f_{+} (f_{-} (g_{-}-g_{+})-g_{+}^2 x_3) \phi_0(z)  \phi_P^A(x_3)
+2 r_{03} \sqrt{\eta } (f_{+} (\phi_P^P(x_3)-\phi_P^T(x_3))\non
&+&(f_{-}+g_{+} x_3 ) (\phi_P^P(x_3)+\phi_P^T(x_3)))\phi_s(z)\big] E_a(t_f^V)\; h^V_f(\alpha^V_a,\beta^V_f,b,b_3)S_t(|f_{-}+x_3g_{+}|)
\bigg\},\label{eq:fall01}
\eeq
\beq
F^{LR}_{aV}&=&-F^{LL}_{aV},\label{eq:falr01}
\eeq
\beq
F^{SP}_{aV}&=&16\pi C_F m^4_B f_B/\sqrt{6} \int_0^1 dz dx_3 \int_0^\infty b db\; b_3 db_3\;\non
&\times&\bigg\{\big[
(2 r_{03} (f_{+}( g_{+}+f_{-} z)-f_{-} g_{-}) \sqrt{\eta } \phi_0(z)  \phi_P^P(x_3)
-(f_{-} g_{-}+g_{+} (f_{+}(1-z)+2 g_{-})) \eta  \phi_s(z)\phi_P^A(x_3)\non
&+&f_{-} f_{+} (f_{-} g_{-}+f_{+} g_{+} (z-1)) \phi_t(z)\phi_P^A(x_3))/\sqrt{\eta}
\big] \non
&\times&E_a(t_e^V)\; h^V_e(\alpha^V_a,\beta^V_e,b,b_3)S_t(z)\non
&+&\big[
f_{+} r_{03} \phi_0(z)  (g_{+} x_3 (\phi_P^P(x_3)-\phi_P^T(x_3))-2 f_{-} \phi_P^T(x_3))
-2 (f_{-} g_{-}+g_{+} (f_{+}+g_{-} x_3)) \sqrt{\eta }\phi_s(z)\phi_P^A(x_3)
\big]\non
&\times& \; E_a(t_f^V)\; h^V_f(\alpha^V_a,\beta^V_f,b,b_3)S_t(|f_{-}+x_3g_{+}|)
\bigg\},\label{eq:fasp01}
\eeq
\beq
M^{LL}_{aV}&=&32\pi C_F m^4_B/\sqrt{6}/\sqrt{\eta} \int_0^1 dx_B dz dx_3 \int_0^\infty b_B db_B\; b db\; \phi_B(x_B,b_B)\non
&\times&\bigg\{\big[
(f_{+} g_{+}+f_{-} g_{-} (g_{-}+g_{+}-1)+f_{+} (g_{-}+g_{+}) (g_{+} (x_3-1)+x_B)\non
&+&f_{-} f_{+} (g_{-}+g_{+}) z) \sqrt{\eta } \phi_0(z)  \phi_P^A(x_3)
+r_{03} \eta  ((g_{-}+g_{+}(1- x_3)-x_B+f_{+} z-4) \phi_P^P(x_3)\non
&-&(g_{-}+g_{+} (x_3-1)+x_B+f_{+} z) \phi_P^T(x_3))\phi_s(z)
+f_{-} f_{+} r_{03} ((f_{+} z+g_{-}) (\phi_P^P(x_3)-\phi_P^T(x_3))\non
&+&(x_B+g_{+} (x_3-1)) (\phi_P^P(x_3)+\phi_P^T(x_3))) \phi_t(z)
\big]E_n(t_g^V)\; h^V_g(\alpha^V_a,\beta^V_g,b_B,b)\non
&+&\big[
(f_{-}-f_{+}) (f_{+} g_{+} (1-z)+g_{-} (f_{-}-x_B+g_{+} x_3)) \sqrt{\eta } \phi_0(z)  \phi_P^A(x_3)\non
&+&(r_{03} \eta\phi_s(z)+f_{-} f_{+} r_{03}\phi_t(z)) (f_{-}+g_{+}x_3-x_B)(\phi_P^P(x_3)-\phi_P^T(x_3))\non
&+&(f_{-} f_{+} r_{03}\phi_t(z)-r_{03} \eta\phi_s(z))f_{+}(z-1) (\phi_P^P(x_3)+\phi_P^T(x_3))\big]\non
&\times& E_n(t_h^V)\; h^V_h(\alpha^V_a,\beta^V_h,b_B,b)
\bigg\},\label{eq:mall02}
\eeq
\beq
M^{LR}_{aV}&=&-32\pi C_F m^4_B/\sqrt{6}/\sqrt{\eta} \int_0^1 dx_B dz dx_3 \int_0^\infty b_B db_B\; b db \; \phi_B(x_B,b_B)\non
&\times&\bigg\{\big[
r_{03} \sqrt{\eta } \phi_0(z)  (f_{-} (1+g_{-}+f_{+} z) (\phi_P^P(x_3)-\phi_P^T(x_3))+f_{+} (g_{+} (x_3-1)-1\non
&+&x_B) (\phi_P^P(x_3)+\phi_P^T(x_3)))
+(g_{-} (g_{+}(x_3-2)-1+x_B)-g_{+} (1+f_{+} z)) \eta  \phi_P^A(x_3)\phi_s(z)\non
&+&f_{-} f_{+} (g_{+}+g_{-} (g_{+} x_3+x_B-1)
+f_{+} g_{+} z) \phi_P^A(x_3) \phi_t(z)\big] E_n(t_g^V)\; h^V_g(\alpha^V_a,\beta^V_g,b_B,b)\non
&-&\big[
f_{+} r_{03} \sqrt{\eta } \phi_0(z)  ((g_{+} x_3-x_B) (\phi_P^P(x_3)+\phi_P^T(x_3))+f_{-} (z \phi_P^P(x_3)+(2-z) \phi_P^T(x_3)))\non
&+&(f_{+} g_{+}(1-z)+g_{-} (f_{-}+g_{+} x_3-x_B)) \eta  \phi_P^A(x_3)\phi_s(z)
+f_{-} f_{+} (g_{-} (f_{-}+g_{+} x_3\non
&-&x_B)+f_{+} g_{+} (-1+z)) \phi_P^A(x_3) \phi_t(z)\big]E_n(t_h^V)\; h^V_h(\alpha^V_a,\beta^V_h,b_B,b)
\bigg\},\label{eq:malr02}
\eeq
\beq
M^{SP}_{aV}&=&32\pi C_F m^4_B/\sqrt{6}/\sqrt{\eta} \int_0^1 dx_B dz dx_3 \int_0^\infty b_B db_B\; b db \; \phi_B(x_B,b_B)\non
&\times&\bigg\{\big[
(g_{-} (f_{-} (1+g_{+} (x_3-2)+x_B)-f_{+} (g_{+} (x_3-2)+x_B))\non
&+&f_{+}g_{+}( (f_{+}-f_{-}) z-1)) \sqrt{\eta } \phi_0(z)  \phi_P^A(x_3)
-r_{03} \eta  ((g_{-}+g_{+}(1- x_3)-x_B+f_{+} z-4) \phi_P^P(x_3)\non
&+&(g_{-}+g_{+} (x_3-1)+x_B+f_{+} z) \phi_P^T(x_3))\phi_s(z)
+f_{-} f_{+} r_{03} ((g_{-}+g_{+} (x_3-1)+x_B\non
&+&f_{+} z) \phi_P^P(x_3)+(g_{-}+g_{+}(1- x_3)-x_B+f_{+} z) \phi_P^T(x_3)) \phi_t(z)
\big] \; E_n(t_g^V)\; h^V_g(\alpha^V_a,\beta^V_g,b_B,b)\non
&+&\big[
f_{+} (g_{-}+g_{+}) (g_{+} x_3-x_B+f_{-} z) \sqrt{\eta } \phi_0(z)  \phi_P^A(x_3)
+(r_{03}\eta\phi_s(z)+f_{-} f_{+} r_{03}\phi_t(z))f_{+} (z-1)\non
&\times&(\phi_P^P(x_3)-\phi_P^T(x_3))+(f_{-} f_{+} r_{03}\phi_t(z)-r_{03}\eta\phi_s(z)) (g_{+} x_3-x_B+f_{-}) (\phi_P^P(x_3)+\phi_P^T(x_3))\big]\non
&\times&E_n(t_h^V)\; h^V_h(\alpha^V_a,\beta^V_h,b_B,b)
\bigg\},\label{eq:masp02}
\eeq
%\item[$\bullet$]  $B_{(s)} \to VP$ decay modes (The meson $P$ is moving along the direction of
%$n$ and resonance $V$ is along the direction of $v$):
\beq
F^{LL}_{eP}&=&8 \pi C_F  m_B^4 F(w^2) \int_0^1 dx_B dx_3 \int_0^\infty b_B\; db_B\; b_3\; db_3\; \phi_B(x_B,b_B)\non
&\times&\bigg\{\big[
(f_{-} g_{-}-f_{+} g_{+} (1+g_{+} x_3)) \phi_P^A(x_3)
+r_{03} (f_{-}+f_{+} (2 g_{+} x_3-1)) \phi_P^P(x_3)\non
&-&r_{03} (f_{-}+f_{+}-2 f_{+} g_{+} x_3) \phi_P^T(x_3)
\big]E_e(t_a^P)\; h^P_a(\alpha^P_e,\beta^P_a,b_B,b_3)S_t(x_3)\non
&-&\big[
-g_{+} (f_{+} g_{-}+f_{-} (x_B-g_{-})) \phi_P^A(x_3)
+2 r_{03} (f_{+} g_{+}+f_{-} (x_B-g_{-})) \phi_P^P(x_3)
\big]\non
&\times&  \; E_e(t_b^P)\; h^P_b(\alpha^P_e,\beta^P_b,b_B,b_3)S_t(|x_B-g_{-}|)\bigg\},\label{eq:fll02}
\eeq
\beq
F^{LR}_{eP}&=&F^{LL}_{eP},  \label{eq:flr02}
\eeq
\beq
F^{SP}_{eP}&=&0,
\eeq
\beq
M^{LL}_{eP}&=&32\pi C_F m^4_B/\sqrt{6} \int_0^1 dx_B dz dx_3 \int_0^\infty b_B db_B\; b db\; \phi_B(x_B,b_B) \phi_0(z) \non
&\times&\bigg\{\big[
(f_{-}-f_{+}) (f_{-} g_{-}+g_{+} (f_{+}(1-z)+g_{-} x_3-x_B)) \phi_P^A(x_3)
+r_{03} (f_{+} g_{+} x_3+f_{-} x_B\non
&+&f_{-} f_{+} z) \phi^P_P+r_{03} (-f_{+} g_{+} x_3+f_{-} (x_B+f_{+} (z-2))) \phi_P^T(x_3)
 \big]  \; E_n(t_c^P)\; h^P_c(\alpha^P_e,\beta^P_c,b_B,b)\non
&-&\big[
(f_{-} g_{-}-f_{+} g_{+}) (f_{+}z+g_{+}x_3-x_B) \phi_P^A(x_3)
+r_{03} (f_{+} g_{+} x_3+f_{-} (x_B-f_{+} z) \phi_P^P(x_3)\non
&+&r_{03} (f_{+} g_{+} x_3-f_{-} (x_B-f_{+}z))) \phi_P^T(x_3)
\big] \; E_n(t_d^P)\; h^P_d(\alpha^P_e,\beta^P_d,b_B,b)
\bigg\},\label{eq:mll02}
\eeq
\beq
M^{LR}_{eP}&=&32\pi C_F m^4_B /\sqrt{6}/\sqrt{\eta} \int_0^1 dx_B dz dx_3 \int_0^\infty b_B db_B\; b db\; \phi_B(x_B,b_B)\non
&\times& \bigg\{\big[
-(f_{-} g_{-}+g_{+} (f_{+}+g_{-} x_3-x_B-f_{+} z)) \eta  \phi_P^A(x_3) \phi_s(z)+f_{-} f_{+} (f_{-} g_{-}+g_{+} (g_{-} x_3+x_B\non
&+&f_{+} (-1+z))) \phi_P^A(x_3) \phi_t(z)
+r_{03} \phi_P^P(x_3) (-(f_{-}+f_{+}(1-z)+g_{+} x_3-x_B) \eta  \phi_s(z)+f_{-} f_{+} (f_{-}+g_{+}x_3\non
&+&x_B+f_{+} (-1+z)) \phi_t(z))
+r_{03} \phi_P^T(x_3) (-(f_{-}+g_{+} x_3+x_B+f_{+} (-1+z)) \eta  \phi_s(z)\non
&+&f_{-} f_{+} (f_{-}+f_{+}(1-z)+g_{+} x_3-x_B) \phi_t(z))
\big]  \; E_n(t_c^P)\; h^P_c(\alpha^P_e,\beta^P_c,b_B,b)\non
&+&\big[
g_{+} \phi_P^A(x_3) ((g_{-} x_3-x_B+f_{+} z) \eta  \phi_s(z)+f_{-} f_{+} (g_{-} x_3+x_B-f_{+} z) \phi_t(z))\non
&+&r_{03} \phi_P^P(x_3) ((g_{+} x_3-x_B+f_{+} z) \eta  \phi_s(z)+f_{-} f_{+} (g_{+} x_3+x_B-f_{+} z) \phi_t(z))\non
&+&r_{03} \phi_P^T(x_3) ((g_{+} x_3+x_B-f_{+} z) \eta  \phi_s(z)+f_{-} f_{+} (g_{+} x_3-x_B+f_{+} z) \phi_t(z))\big]\non
&\times&\;E_n(t_d^P)\; h^P_d(\alpha^P_e,\beta^P_d,b_B,b)
\bigg\},\label{eq:mlr02}
\eeq
\beq
M^{SP}_{eP}&=&32\pi C_F m^4_B/\sqrt{6}  \int_0^1 dx_B dz dx_3 \int_0^\infty b_B db_B\; b db\; \phi_B(x_B,b_B) \phi_0(z)\non
&\times&\bigg\{\big[
(f_{-} g_{-}-f_{+} g_{+}) (f_{-}+f_{+}(1-z)+g_{+} x_3-x_B) \phi_P^A(x_3)
+r_{03} (f_{+} g_{+} x_3+f_{-} x_B\non
&+&f_{-} f_{+} z) \phi_P^P(x_3)+r_{03} (f_{+} g_{+} x_3-f_{-} (x_B+f_{+} (z-2))) \phi_P^T(x_3)
\big] \; E_n(t_c^P)\; h^P_c(\alpha^P_e,\beta^P_c,b_B,b)\non
&-&\big[
(f_{-}-f_{+}) g_{+} (g_{-} x_3-x_B+f_{+} z) \phi_P^A(x_3)
+r_{03} (f_{+} g_{+} x_3+f_{-} x_B-f_{-} f_{+} z) \phi_P^P(x_3)\non
&-&r_{03} (f_{+} g_{+} x_3+f_{-}( f_{+}z-x_B)) \phi_P^T(x_3)
\big] E_n(t_d^P)\; h^P_d(\alpha^P_e,\beta^P_d,b_B,b)
\bigg\},\label{eq:msp02}
\eeq
\beq
F^{LL}_{aP}&=&-8\pi C_F m^4_B f_B \int_0^1 dz dx_3 \int_0^\infty b db\; b_3 db_3\;\non
&\times&\bigg\{\big[
(f_{-} g_{-}+f_{+} g_{+} (g_{+} x_3-1)) \phi_0(z) \phi_P^A(x_3)
+2 r_{03} (g_{+} x_3-2) \sqrt{\eta } \phi_P^P(x_3) \phi_s(z)
-2 g_{+} r_{03} x_3 \sqrt{\eta } \phi_P^T(x_3) \phi_s(z) \big]\non
&\times&\; E_a(t_e^P)\; h^P_e(\alpha^P_a,\beta^P_e,b,b_3)S_t(x_3)\non
&+&\big[
g_{+} (f_{+} (g_{-}+f_{+} z)-f_{-} g_{-}) \phi_0(z)  \phi_P^A(x_3)
+2 r_{03} \phi_P^P(x_3) ((g_{-}+g_{+}+f_{+} z) \sqrt{\eta } \phi_s(z)\non
&+&f_{-} f_{+} (g_{-}-g_{+}+f_{+} z) \phi_t(z)/\sqrt{\eta })
 \big]\; E_a(t_f^P)\; h^P_f(\alpha^P_a,\beta^P_f,b,b_3)S_t(|g_{-}+zf_{+}|)
\bigg\},\label{eq:fall02}
\eeq
\beq
F^{LR}_{aP}&=&-F^{LL}_{aP},\label{eq:falr02}
\eeq
\beq
F^{SP}_{aP}&=&16\pi C_F m^4_B f_B\int_0^1 dz dx_3 \int_0^\infty b db\; b_3 db_3\;\non
&\times&\bigg\{\big[
2 (g_{-} (g_{+} x_3-1)-g_{+}) \sqrt{\eta } \phi_P^A(x_3) \phi_s(z)
+r_{03} (f_{-}+f_{+} (g_{+} x_3-1)) \phi_0(z)  \phi_P^P(x_3)
-r_{03} (f_{-}+f_{+}\non
&-&f_{+} g_{+} x_3) \phi_0(z)  \phi_P^T(x_3)
\big] \; E_a(t_e^P)\;h^P_e(\alpha^P_a,\beta^P_e,b,b_3)S_t(x_3)\non
&+&\big[\phi_P^A(x_3) (-2 g_{-} g_{+} \eta  \phi_s(z)+f_{+} g_{+} z (f_{-} f_{+} \phi_t(z)-\eta  \phi_s(z)))/\sqrt{\eta }
+2 r_{03} (-f_{+} g_{+}+f_{-} (g_{-}+f_{+} z)) \phi_0(z)  \phi_P^P(x_3)
\big]\non
&\times&\; E_a(t_f^P)\; h^P_f(\alpha^P_a,\beta^P_f,b,b_3)S_t(|g_{-}+zf_{+}|)
\bigg\},\label{eq:fasp02}
\eeq
\beq
M^{LL}_{aP}&=&32\pi C_F m^4_B/\sqrt{6}/\sqrt{\eta} \int_0^1 dx_B dz dx_3 \int_0^\infty b_B db_B\; b db\; \phi_B(x_B,b_B)\non
&\times&\bigg\{\big[
(f_{-}^2 g_{-}-f_{-} (g_{-} (1+f_{+}-g_{+} x_3)+g_{+} (x_B+f_{+} (z-1)))+f_{+} g_{+} (1-g_{-} x_3+x_B\non
&+&f_{+} (z-1))) \sqrt{\eta } \phi_0(z)  \phi_P^A(x_3)
+r_{03} \phi_P^P(x_3) (-(-4+f_{-}+f_{+}(1-z)+g_{+} x_3-x_B) \eta  \phi_s(z)+f_{-} f_{+} (f_{-}\non
&+&g_{+} x_3+x_B+f_{+} (z-1)) \phi_t(z))
+r_{03} \phi_P^T(x_3) (-(f_{-}+g_{+} x_3+x_B+f_{+} (z-1)) \eta  \phi_s(z)\non
&+&f_{-} f_{+} (f_{-}+f_{+}(1-z)+g_{+} x_3-x_B) \phi_t(z))
\big]
\; E_n(t_g^P)\; h^P_g(\alpha^P_a,\beta^P_g,b_B,b)\non
&+&\big[
(g_{-}+g_{+}) (f_{+} g_{+} (x_3-1)+f_{-} (g_{-}-x_B+f_{+} z)) \sqrt{\eta } \phi_0(z)  \phi_P^A(x_3)
+r_{03} \phi_P^P(x_3) (-(g_{-}+g_{+}(1-x_3)\non
&-&x_B+f_{+} z) \eta  \phi_s(z)+f_{-} f_{+} (g_{-}+g_{+} (x_3-1)-x_B+f_{+} z) \phi_t(z))
+r_{03} \phi_P^T(x_3) ((-g_{-}+g_{+}(1-x_3)\non
&+&x_B-f_{+} z) \eta  \phi_s(z)+f_{-} f_{+} (g_{-}+g_{+}(1-x_3)-x_B+f_{+} z) \phi_t(z))
\big]\non
&\times& \; E_n(t_h^P)\; h^P_h(\alpha^P_a,\beta^P_h,b_B,b)
\bigg\},\label{eq:all02}
\eeq
\beq
M^{LR}_{aP}&=&32\pi C_F m^4_B/\sqrt{6}/\sqrt{\eta} \int_0^1 dx_B dz dx_3 \int_0^\infty b_B db_B\; b db \; \phi_B(x_B,b_B)\non
&\times&\bigg\{\big[\phi_P^A(x_3) (g_{-} (1+f_{-}+g_{+} x_3) (f_{-} f_{+} \phi_t(z)-\eta  \phi_s(z))+g_{+} (-1+x_B+f_{+} (z-1)) (\eta  \phi_s(z)\non
&+&f_{-} f_{+} \phi_t(z)))+r_{03} (f_{+}(1+g_{+} x_3)+f_{-} (x_B+f_{+} z-1)) \sqrt{\eta } \phi_0(z)  \phi_P^P(x_3)+r_{03} (-f_{+} (1\non
&+&g_{+} x_3)+f_{-} (-1+x_B+f_{+} (z-2))) \sqrt{\eta } \phi_0(z)  \phi_P^T(x_3)
\big]  E_n(t_g^P)\; h^P_g(\alpha^P_a,\beta^P_g,b_B,b)\non
&-&\big[g_{+} \phi_P^A(x_3) (-(g_{-} (x_3-2)+x_B-f_{+} z) \eta  \phi_s(z)+f_{-} f_{+} (g_{-} x_3-x_B+f_{+} z) \phi_t(z))\non
&+&r_{03} (f_{+} g_{+} (x_3-1)+f_{-} (g_{-}-x_B+f_{+} z)) \sqrt{\eta } \phi_0(z)  \phi_P^P(x_3)
+r_{03} (-f_{+} g_{+} (x_3-1)\non
&+&f_{-} (g_{-}-x_B+f_{+} z)) \sqrt{\eta } \phi_0(z)  \phi_P^T(x_3)
\big] E_n(t_h^P)\; h^P_h(\alpha^P_a,\beta^P_h,b_B,b)
\bigg\},\label{eq:alr02}
\eeq
\beq
M^{SP}_{aP}&=&32\pi C_F m^4_B/\sqrt{6}/\sqrt{\eta} \int_0^1 dx_B dz dx_3 \int_0^\infty b_B db_B\; b db \; \phi_B(x_B,b_B)\non
&\times&\bigg\{\big[
(f_{+} g_{+} (-1+(g_{-}+g_{+}) x_3)+f_{-} (g_{-}(1+x_B)+g_{+} x_B+f_{+} (g_{-}+g_{+}) z)) \sqrt{\eta } \phi_0(z)  \phi_P^A(x_3)\non
&+&r_{03} \phi_P^P(x_3) ((-4+f_{-}+f_{+}(1-z)+g_{+} x_3-x_B) \eta  \phi_s(z)+f_{-} f_{+} (f_{-}+g_{+} x_3+x_B+f_{+} (-1\non
&+&z)) \phi_t(z))+r_{03} \phi_P^T(x_3) (-(f_{-}+g_{+} x_3+x_B+f_{+} (z-1)) \eta  \phi_s(z)-f_{-} f_{+} (f_{-}+f_{+}(1-z)\non
&+&g_{+} x_3-x_B) \phi_t(z))\big]E_n(t_g^P)\; h^P_g(\alpha^P_a,\beta^P_g,b_B,b)\non
&+&\big[
(f_{-}-f_{+}) g_{+} (g_{-} (x_3-2)+x_B-f_{+} z) \sqrt{\eta } \phi_0(z)  \phi_P^A(x_3)
+r_{03} \phi_P^P(x_3) ((g_{-}+g_{+}(1-x_3)-x_B\non
&+&f_{+} z) \eta  \phi_s(z)+f_{-} f_{+} (g_{-}+g_{+} (x_3-1)-x_B+f_{+} z) \phi_t(z))
+r_{03} \phi_P^T(x_3) ((-g_{-}+g_{+}(1- x_3)\non
&+&x_B-f_{+} z) \eta  \phi_s(z)-f_{-} f_{+} (g_{-}+g_{+}(1-x_3)-x_B+f_{+} z) \phi_t(z))
\big]E_n(t_h^P)\; h^P_h(\alpha^P_a,\beta^P_h,b_B,b)
\bigg\},\label{eq:asp02}
\eeq
with the color factor $C_F=4/3$ and the mass ratio $r_{03}=m_{03}/m_{B_{(s)}}$.
%\end{enumerate}

The evolution factors $E_i(t)$, $i=e,a,n$, in the above factorization formulas are written as
\begin{eqnarray}
E_e(t)&=&\alpha_s(t) \exp[-S_B(t)-S_V(t)],\nonumber\\
E_a(t)&=&\alpha_s(t) \exp[-S_P(t)-S_V(t)],\nonumber\\
E_n(t)&=&\alpha_s(t) \exp[-S_B(t)-S_V(t)-S_P(t)],
\end{eqnarray}
where the Sudakov exponents $S_{B,P,V}$ are given by
\begin{eqnarray}
S_B&=& s(x_B\frac{m_B}{\sqrt2},b_B )+\frac53\int^t_{1/b_B}\frac{d\bar\mu}{\bar\mu} \gamma_q(\alpha_s(\bar\mu)),\nonumber\\
S_V&=& s(\frac{m_B}{\sqrt2} zf_+,b )+ s(\frac{m_B}{\sqrt2}(1-z)f_+,b )
+ 2\int^t_{1/b}\frac{d\bar\mu}{\bar\mu} \gamma_q(\alpha_s(\bar\mu)),\nonumber\\
S_P&=& s(\frac{m_B}{\sqrt2}x_3g_+,b_3 ) +s(\frac{m_B}{\sqrt2}(1-x_3)g_+,b_3 )
+2\int^t_{1/b_3}\frac{d\bar\mu}{\bar\mu} \gamma_q(\alpha_s(\bar\mu)),
\end{eqnarray}
with the quark anomalous dimension $\gamma_q=-\alpha_s/\pi$.
The explicit expressions of the functions $s(Q,b)$ can be found in the Appendix of Ref.~\cite{prd76-074018}.

The threshold resummation factor $S_t(x)$ takes the form
\begin{eqnarray}
\label{eq-def-stx}
S_t(x)=\frac{2^{1+2c}\Gamma(3/2+c)}{\sqrt{\pi}\Gamma(1+c)}[x(1-x)]^c,
\end{eqnarray}
where $c=0.3$ is adopted in our numerical analysis.

The hard functions $h^{V(P)}_i$, $i=a$-$h$, in the factorization formulas are written as
\begin{eqnarray}
h^{V(P)}_i(\alpha^{V(P)}_j,\beta^{V(P)}_i,b_1,b_2)&=&h_1(\alpha^{V(P)}_j,b_1) h_2(\beta^{V(P)}_i,b_1,b_2),\quad j=e,a,\nonumber\\
h_1(\alpha^{V(P)}_j,b_1)&=&\left\{\begin{array}{ll}
K_0(\sqrt{\alpha^{V(P)}_j} b_1), & \quad  \quad \alpha^{V(P)}_j >0\\
K_0(i\sqrt{-\alpha^{V(P)}_j} b_1),& \quad  \quad \alpha^{V(P)}_j<0
\end{array} \right.\nonumber\\
h_2(\beta^{V(P)}_i,b_1,b_2)&=&\left\{\begin{array}{ll}
\theta(b_1-b_2)I_0(\sqrt{\beta^{V(P)}_i}b_2)K_0(\sqrt{\beta^{V(P)}_i}b_1)+(b_1\leftrightarrow b_2), & \quad   \beta^{V(P)}_i >0\\
\theta(b_1-b_2)I_0(\sqrt{-\beta^{V(P)}_i}b_2)K_0(i\sqrt{-\beta^{V(P)}_i}b_1)+(b_1\leftrightarrow b_2),& \quad   \beta^{V(P)}_i<0
\end{array} \right.
\end{eqnarray}
with the Bessel function $K_0(ix)=\pi[-N_0(x)+i J_0(x)]/2$, and the virtuality $\alpha^{V(P)}_j$
($\beta^{V(P)}_i$) of the internal gluon (quark) in the diagrams:
\begin{eqnarray}
\alpha^V_e&=&f_{+}zx_B,\quad\quad \alpha^V_a=f_{+}(z-1)(f_{-}+g_{+}x_3),\non
\beta^V_a&=&zf_{+},~\quad\quad\quad  \beta^V_b=(x_B-f_{-})f_{+}\non
\beta^V_c&=&(g_{-}+f_{+}z)(x_B+g_{+}(x_3-1)), \quad\quad \quad\quad\beta^V_d=f_{+}z(x_B-g_{+}x_3),\non
\beta^V_e&=&-(f_{-}+g_{+})(f_{+}(1-z)+g_{-}),~\,\quad\quad \quad\quad\beta^V_f=-(f_{-}+g_{+}x_3)f_{+},\non
\beta^V_g&=&1+(g_{-}+f_{+}z)(x_B+g_{+}(x_3-1),~~\,\quad\quad \beta^V_h=f_{+}(z-1)(f_{-}-x_B+g_{+}x_3),
\end{eqnarray}
and
\begin{eqnarray}
\alpha^P_e&=&g_{+}x_3x_B,\quad\quad\quad \alpha^P_a=g_{+}(x_3-1)(g_{-}+f_{+}z),\non
\beta^P_a&=&x_3g_{+},~\quad\quad\quad\quad  \beta^P_b=(x_B-g_{-})g_{+}\non
\beta^P_c&=&(f_{-}+g_{+}x_3)(x_B+f_{+}(z-1)), \quad\quad \quad\quad\beta^P_d=g_{+}x_3(x_B-f_{+}z),\non
\beta^P_e&=&x_3g_{+}-1,\quad\quad \quad\beta^P_f=-(g_{-}+f_{+}z)g_{+},\non
\beta^P_g&=&1+(f_{-}+g_{+}x_3)(x_B+f_{+}(z-1),~~\,\quad\quad \beta^P_h=g_{+}(x_3-1)(g_{-}-x_B+f_{+}z).
\end{eqnarray}

The hard scales $t^{V(P)}_i$, $i=a$-$h$, are chosen as the maxima of the virtualities
involved in the decays described by Fig.~\ref{fig:fig1} (Fig.~\ref{fig:fig2}):
\begin{eqnarray}
t^V_{a,b}&=&\max\{m_B\sqrt{|\alpha^V_e|},m_B\sqrt{|\beta^V_{a,b}|},1/b,1/b_B\},\non
t^V_{c,d}&=&\max\{m_B\sqrt{|\alpha^V_e|},m_B\sqrt{|\beta^V_{c,d}|},1/b_3,1/b_B\},\non
t^V_{e,f}&=&\max\{m_B\sqrt{|\alpha^V_a|},m_B\sqrt{|\beta^V_{e,f}|},1/b,1/b_3\},\non
t^V_{g,h}&=&\max\{m_B\sqrt{|\alpha^V_a|},m_B\sqrt{|\beta^V_{g,h}|},1/b,1/b_B\},
\end{eqnarray}
and
\begin{eqnarray}
t^P_{a,b}&=&\max\{m_B\sqrt{|\alpha^P_e|},m_B\sqrt{|\beta^P_{a,b}|},1/b_3,1/b_B\},\non
t^P_{c,d}&=&\max\{m_B\sqrt{|\alpha^P_e|},m_B\sqrt{|\beta^P_{c,d}|},1/b,1/b_B\},\non
t^P_{e,f}&=&\max\{m_B\sqrt{|\alpha^P_a|},m_B\sqrt{|\beta^P_{e,f}|},1/b,1/b_3\},\non
t^P_{g,h}&=&\max\{m_B\sqrt{|\alpha^P_a|},m_B\sqrt{|\beta^P_{g,h}|},1/b,1/b_B\}.
\end{eqnarray}

%=============== Refs ===============%

\end{document}